\documentclass[10pt,superscriptaddress,twocolumn,amsmath,amssymb,aps,prl,showpacs]{revtex4}
\usepackage{mathrsfs}
\usepackage{graphicx}% Include figure files
\usepackage{dcolumn}% Align table columns on decimal point
\usepackage{bm}
\usepackage{amssymb}
\usepackage{amsmath}
\usepackage{paralist}
\usepackage{color}

\newcommand{\Tr}{\mathrm{Tr}}
\def\be{\begin{equation}}
\def\ee{\end{equation}}
\def\bea{\begin{eqnarray}}
\def\eea{\end{eqnarray}}

\newif\ifNOSUP \NOSUPfalse

\begin{document}

\title{Exact quantum dynamics of  XXZ central spin problems }

\author{Wen-Bin He}
\affiliation{State Key Laboratory of Magnetic Resonance and Atomic and Molecular Physics,
Wuhan Institute of Physics and Mathematics, Chinese Academy of Sciences, Wuhan 430071, China}
\affiliation{University of Chinese Academy of Sciences, Beijing 100049, China}

\author{Stefano Chesi}
\email[]{stefano.chesi@csrc.ac.cn}
\affiliation{Beijing Computational Science Research Center, Beijing 100193, China}

\author{Hai-Qing  Lin}
%\email[e-mail:]{haiqing0@csrc.ac.cn}
\affiliation{Beijing Computational Science Research Center, Beijing 100193, China}

\author{Xi-Wen Guan}
\email[]{xwe105@wipm.ac.cn}
\affiliation{State Key Laboratory of Magnetic Resonance and Atomic and Molecular Physics,
Wuhan Institute of Physics and Mathematics, Chinese Academy of Sciences, Wuhan 430071, China}
\affiliation{Center for Cold Atom Physics, Chinese Academy of Sciences, Wuhan 430071, China}
\affiliation{Department of Theoretical Physics, Research School of Physics and Engineering,
Australian National University, Canberra ACT 0200, Australia}

\date{\today}


\begin{abstract}
We obtain analytically  close forms of benchmark quantum dynamics of the collapse and revival  (CR), reduced density matrix, Von Neumann  entropy,  and fidelity  for the XXZ central spin problem. These quantities characterize  the quantum decoherence and entanglement of the system with few to many bath spins,  and for  a short to infinitely long time evolution. 
For the homogeneous central spin problem,  the effective magnetic field  $B$,  coupling constant  $A$  and longitudinal interaction $\Delta$  significantly  influence the time scales of the quantum dynamics of the central spin and the bath, providing  a  tunable resource  for quantum metrology. 
%
%{\color{red}The presence of a finite longitudinal interaction  $\Delta$   allows for  quantum revivals even at a very  small number of bath spins $N$,   facilitating experimental control of  entangled states.  }
%
Under  the resonance condition  $B=\Delta=A$, the location of the $m$-th revival peak in time reaches  a simple relation 
$t_{r} \simeq\frac{\pi N}{A} m$ for a large  $N$. 
For  $\Delta =0$,   $N\to \infty$ and  a small  polarization in  the initial spin coherent state,  our analytical  result  for the CR  recovers  the known  expression  found   in the Jaynes-Cummings model,  thus  building up  an exact dynamical connection  between the central spin problems  and   the light-matter interacting systems in quantum nonlinear optics. 
%and  revealing the statistical nature of Holstein-Primakoff transformation.
In addition, the CR dynamics is robust to a moderate inhomogeneity of the coupling amplitudes, while disappearing at strong inhomogeneity.
\end{abstract}
\maketitle

%%%%%%%%%%%%%%%%%%%%%%%%%%%%%%%%%%%%%%%%%

Quantum dynamics of many-body systems has been a long-standing  challenge in physics~\cite{Scully:1997,APolkRMP,Suter:2016,Zurek85,YangW:2017}. 
It is  always a formidable task  for physicist, due to the difficulty of  analytically deriving many-body  eigenfunctions  and  the exponentially growing complexity of numerics \cite{Coish:2010,Faribault:2013,Lindoy:2018}.
Over a decade,  important  progresses, which  have been made in a variety of fields, such as  atomic qubits  coupled to a cavity \cite{Eberly:1980,Gea-Banacloche:1990,Rempe:1987,Shore:1993}, central spin problems \cite{Orany:2011,Dooley:2013,Lindoy:2018,Quan:2006,Chesi:2012,Chesi:2015,Wu:2014,Wu:2016,Wu:2018}, resonant superconductor qubits \cite{Blais:2004,Neill:2016,haohWang:2017,haohWang:2018,JQYou:2013}, long range interacting spin chains of Rydberg atoms \cite{Saffman:2010,DBarredo:2015}, are  greatly improving our understanding of quantum dynamics and entanglement of many-body systems. 

In this context,  exact Bethe ansatz solvable models have been particularly  fruitful to the study of quantum dynamics of this kind, for example, integrability-based central spin problems \cite{Amico:2001,Amico:2001-PRL,Amico:2002,H.QZhou:2002,Yang:2004,JDukelsky:2004,Khaetskii:2002,Khaetskii:2003,Bortz:2007,Bortz:2010,Claeys:2015a,Claeys:2015b}, atom-field interacting systems in quantum nonlinear optics \cite{Scully:1997,Bogoliubov,suyi}, thermalization  and quantum dynamics \cite{Kinoshita2006,Hofferberth:2007,Ronzheimer:2013,Claeys:2018}, and  quantum hydrodynamics \cite{Castro-Alvaredo:2016,Bertini:2016}, etc. 
  However,   the problem of the size of the Hilbert space increasing  exponentially with the particle number  still prohibits full  analytical accesses  to the quantum dynamics  at a many-body level. 
 Therefore, it  is  extremely rare  to find  an exact quantum dynamics of  integrable models. 
Here we circumvent the complexity of the usual Bethe ansatz  \cite{Links:2017} and develop analytical approaches to the homogeneous central spin problems, obtaining a full characterization of their quantum dynamics through simple closed-form expressions. We further substantiate the relevance of the analytical results through exact numerical simulations, showing that main predicted features, like the occurrence of collapse and revival (CR) dynamics, are robust to inhomogeneity.

 \textbf{\bf Quantum  collapse and revival. } One class of  integrable systems with long-range interactions,  called  Richardson-Gaudin models \cite{Richardson,Gaudin:1976,Gaudin-book},  has found interesting applications  in various physical problems \cite{Khaetskii:2002,Khaetskii:2003,Dooley:2013,Kucsko:2016}. 
%
%In particular,  the central spin problem has recently attracted growing interest in quantum devices, for example when a central spin is coupled to a nuclear spin bath \cite{Khaetskii:2002,Khaetskii:2003,Dooley:2013,Kucsko:2016}. 
%
The XXZ  central spin problem, i.e. a central spin coupled  to $N$ bath spins, is described by the Hamiltonian 
 \begin{eqnarray}
 H&=&B\mathbf{\bf s} ^{z}_{0}+2\sum_{j=1}^{N}\left[ A_j(\mathbf{\bf s}_{0}^{x}\mathbf{\bf s}_{j}^{x}+\mathbf{\bf s}_{0}^{y} \mathbf{\bf s}_{j}^{y})+\Delta_j \,\mathbf{\bf s}_{0}^{z} \mathbf{\bf s}_{j}^{z} \right],
 \label{Ham-1}
 \end{eqnarray}
 where $B$ is an effective  external magnetic field for the central spin \cite{note-1}, $N$ is the number of spins in the bath, $A_j$ is the transverse coupling amplitude,  and  $\Delta_j$ is the longitudinal interaction. The model (\ref{Ham-1})  is integrable  if $\Delta_j$ and $A_j$ are related through $\Delta_j^2-A_j^2 ={\rm Const.} $, see \cite{Yang:2004,Claeys:2015a,Claeys:2015b}. 
 Although this type of models, e.g.  (\ref{Ham-1}),  were  known as an exactly solvable model  long time ago \cite{Gaudin:1976}, the binomial sets of Bethe ansatz roots $C_{N+1}^M$ impose a big numerical challenge in calculation of quantum dynamics of this model  \cite{Bortz:2007,Bortz:2010,Faribault:2013,Lindoy:2018}.
 Here $M$  is  the number of total down spins in the system. Importance of  Hamiltonian~(\ref{Ham-1})  is  in  its   promising  applications to realistic problems in quantum metrology,  based on Nitrogen Vacancy (NV)  centers \cite{RBLiu:2017}, highly symmetric molecules with  $N$ nuclear spins coupled to  the nuclear spin of a central atom  \cite{YangW:2017,Jones:2009}, etc.

The general central spin problem with non-uniform couplings, for example,  $A_{j}=A \exp(-\alpha(j-1)/N)$ where $\alpha$ is the  inhomogeneity parameter, is integrable but its dynamics is still challenging to analyze.  
We first  analytically solve the dynamical evolution for the homogeneous case, namely $A_{j}=A, \Delta_{j}=\Delta$,  and later analyze the effect of inhomogeneity for the CR. 
%[Stefano] shorteneed the text
The homogenous central spin problem (see, e.g., Refs \cite{Khaetskii:2003,Dooley:2013,Coish:2007}) enables one to derive exact expressions of the  quantum dynamics since bath spins can map onto a  large spin operator $\mathbf{J}=\sum_{j=1}^{N}\mathbf{s}_{j}$.
Thus Eq.~(\ref{Ham-1}) can be rewritten as 
\begin{equation}
H=B\mathbf{\bf s} ^{z}_{0}+A\left(\mathbf{\bf s}_{0}^{+} \mathbf{\bf J}^{-}+\mathbf{\bf s}_{0}^{-} \mathbf{\bf J}^{+}\right)+2\Delta\mathbf{\bf s}_{0}^{z} \mathbf{\bf J}^{z}.
\label{Ham-2}
\end{equation}
Below we  analytically derive  the CR dynamics, reduced density matrix, Von Neumann  entropy, and fidelity, providing an important  benchmark quantum dynamics of this class of models. 
%The presence of a finite longitudinal interaction  $\Delta$   allows for  quantum revivals even at a very  small number of bath spins $N$,   facilitating experimental control of  entangled states.
 %

 %As usual (see, e.g., Refs \cite{Dooley:2013,Coish:2007}), we  choose the spin coherent state as an initial state for the bath $ \vert \Phi_{bath} \rangle=\otimes_{j=1}^{N}[\sin(\theta/2)|\uparrow \rangle_{j}+\cos(\theta/2) | \downarrow\rangle_{j}] $. 
%The  wave function can be formally determined as following in terms of full sets of eigenfunctions $\left\{ \vert \phi_{k} \rangle \right\}$
%\begin{equation}
%\vert \psi(t) \rangle= \sum_{k} \vert \phi_{k} \rangle \langle \phi_{k} \vert \Phi_{0} \rangle  e^{-iE_{k}t}
%\end{equation}
%where $\vert  \Phi_{0}\rangle=\vert \uparrow \rangle_{0}\otimes  \vert \Phi_{bath} \rangle$, central spin is in spin-up state. We develop a  recurrence method to determine the wave function and   the dynamical property of the system.
%

\begin{figure}
  \begin{center}
%\begin{flushleft}
\includegraphics[width=0.45\textwidth]{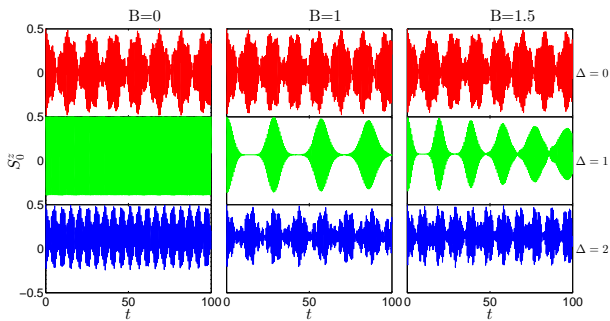}
%\end{flushleft}
\end{center} 
\caption{The  central spin polarization  evolves in time  under  different  values of the magnetic field  $B$,  and longitudinal interaction  $\Delta$. 
In contrast to a two-level  atom coupled to a cavity \cite{Scully:1997}, here we demonstrate that at  resonance $\Delta=A=B=1$, the quantum CR can be observed even for a small system size  $N=8$ (for  $N=4,\,6 $, see  the SM \cite{SuppM}).  Such a small number  of bath spins are experimentally accessible,  for example by superconducting circuits \cite{haohWang:2017,haohWang:2018}.}
\label{Fig1:cr}
\end{figure}

The  phenomenon of quantum CR has long been studied in quantum nonlinear optics ~\cite{Scully:1997}.
However, there  still lacks  a comprehensive  understanding  of such phenomenon in interacting spin systems \cite{Dooley:2013,Khaetskii:2002,Khaetskii:2003,Bortz:2007,Bortz:2010,Tonel:2005}.
 In order  to overcome the exponentially increasing scales  in solving the Bethe ansatz equations of the Gaudin magnet (\ref{Ham-1}),  here we directly calculate   the wave function under a unitary time evolution of the Hamiltonian (\ref{Ham-2}), i.e.  $\vert \psi(t) \rangle= e^{-iHt}\vert \Phi_{0} \rangle$. 
In the initial state  $\vert  \Phi_{0}\rangle=\vert \uparrow \rangle_{0}\otimes  \vert \Phi_{bath} \rangle$ with $ \vert \Phi_{bath} \rangle=\otimes_{j=1}^{N}[\sin(\theta/2)|\uparrow \rangle_{j}+\cos(\theta/2) | \downarrow\rangle_{j}] $ (see, e.g., Refs \cite{Dooley:2013,Coish:2007}). The spin coherent state 
  can be  written in terms of the Dicke states as $\vert \Phi_{bath}\rangle= \sum_{n=0}^{N}\sqrt{C_{N}^{n}}[\sin(\theta/2)]^{n}[\cos(\theta/2)]^{N-n}\vert n \rangle$ with $\vert n \rangle= \vert \frac{N}{2},n- \frac{N}{2}\rangle$. 
 The Dicke state is  the eigenstate of the operators  $\mathbf{J}^{2}$  and $\mathbf{J}^{z}$. For example, $\mathbf{J}^{z}|n\rangle =(-\frac{N}{2}+n)|n\rangle$, $\mathbf{J^-}|n\rangle =\sqrt{b_n} |n-1\rangle$,  and  $\mathbf{J^+}|n\rangle =\sqrt{b_{n+1}} |n+1\rangle$ with $b_n=n(N-n+1)$.
We develop a  recurrence method to determine the wave function and  the dynamical property of the system.
After  a lengthy calculation, we may obtain  an explicit form of  the wave function at arbitrary times, i.e. 
\begin{eqnarray}
\vert  \psi(t) \rangle &=& \sum \limits_{n=0}^{N}\sqrt{C_{N}^{n}}  [\sin(\theta/2)]^{n}[\cos(\theta/2)]^{N-n} \nonumber \\
&&\times  \left[ P_{\downarrow}^{n} (t) \vert \downarrow \rangle_0 \vert n+1\rangle+ P_{\uparrow}^{n}(t) \vert \uparrow \rangle_0 \vert n\rangle\right],
\label{wave-func}
\end{eqnarray}
where the parameters are defined as follows:  $P_{\uparrow}^{n}= -i \left(\Delta_{n+1}/ \Omega_{n+1}\right) \sin( \frac{\Omega_{n+1}t}{2})+\cos( \frac{\Omega_{n+1}t}{2})$ and $P_{\downarrow}^{n}=-i 2\left( \sqrt{b_{n+1}}A/ \Omega_{n+1}\right)\sin( \frac{\Omega_{n+1}t}{2})$ with $\Delta_{n}=B+(2n-1-N)\Delta$ and $\Omega_{n}^2=\Delta^{2}_{n}+4b_nA^2$, see the SM  \cite{SuppM} for a detailed derivation.
 %
%In the following we take the  angle $\theta=\pi/2$ if not specified otherwise. 
%
We observe that the Rabi oscillation frequency $\Omega_n$ has an essential dependence on the coupling  parameters $A$, $\Delta$, and the effective magnetic field $B$. 
The flip-flop interaction, i.e., the second term of Eq.~(\ref{Ham-2}),  leads to the state change between  $ \vert \downarrow \rangle_0\vert n+1\rangle$ and $ \vert \uparrow \rangle_0 \vert n\rangle$.

Using the closed-form of the wave function (\ref{wave-func}),   the  time evolution of the central  spin polarization  $S^{z}_{0}(t)= \langle \psi(t) \vert \mathbf{s}^{z}_{0}\vert \psi(t) \rangle$ can be calculated in a straightforward way. 
By a lengthy algebra, we obtain an  explicit form of the quantum CR of the homogeneous central spin problem
\begin{eqnarray}
S_{0}^{z}(t)&=&\frac{1}{2}\sum_{n=0}^{N}C_{N}^{n}[\sin^{2}(\theta/2)]^{n}[\cos^{2}(\theta/2)]^{N-n}\nonumber \\
&& \times \left[ \frac{ \Delta^{2}_{n+1}}{(\Omega_{n+1})^2}+\frac{4b_{n+1}A^2}{(\Omega_{n+1})^2} \cos(\Omega_{n+1}t) \right]. 
\label{eqnarray}
\end{eqnarray}
This compact form of the quantum CR describes  a very  rich quantum dynamics of the  many-body problems (\ref{Ham-2}), see Fig.~\ref{Fig1:cr}. 
The effective magnetic field $B$, the coupling constant $A$, the longitudinal interaction  $\Delta$,  and the number of bath spins $N$  all play an important role in controlling the features of  the quantum CR dynamics. 
The Rabi oscillation frequency $\Omega_{n}=\sqrt{\Delta^{2}_{n}+4b_nA^2}$  depends  not only on  the  the Dicke state  $n$  but also  on   the coupling constant $A$ and the longitudinal interaction  $\Delta$. 
For a large $\Delta$, the collapses disappear whereas the revival period become shorter. 
An almost perfect CR occurs at  the resonance $\Delta=A=B$. 
The   interaction effect,  driven  by $\Delta$, strongly   influences the frequency of the oscillation, amplitudes of the revivals and   the fidelity of the central spin. 
It is particularly interesting to observe that  a finite $\Delta$  facilitates  the quantum revivals  even for a very  small number of bath spins.
This can help with experimental control of quantum dynamic transfers  in such kind   of systems, see  the SM \cite{SuppM}  for a detailed discussion.

%% the eigenfunction for spin-up 
%\begin{widetext}
\begin{figure}
%\begin{figure}[ht]
  \begin{center}
%\begin{flushleft}
\includegraphics[scale=0.33]{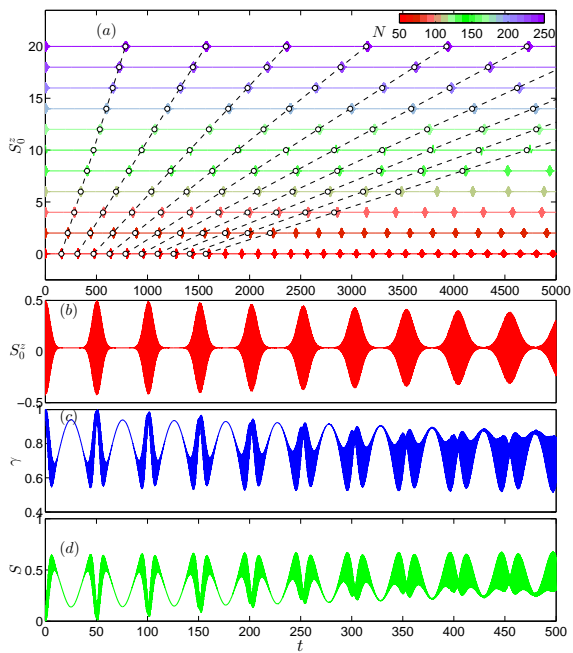}
%\end{flushleft}
\end{center} 
\caption{(a):  Time evolution of the central spin polarization  for  different bath spin sizes $N\in [50-250]$, with particle number step $\delta N=20$.  All the curves are shifted upward  by one  for $N>50$. We use the circles and dashed lines to mark the revival peaks. The times of the revival peaks  depend  linearly  on  the bath size $N$, see the  main  text.  
Time evolution of the central spin polarization for a smaller value of $N=15$ (b), together with the corresponding quantum purity (c), and Von Neumann entropy (d).  Here $B=\Delta=A=1$. }
\label{Fig2:cr-entropy}
%\end{figure} 
\end{figure} 
%\end{widetext}

Under such resonance  condition $B=\Delta=A$, the oscillation frequency becomes $\Omega_{n}=A\sqrt{4n+N^2}$.
 Thus  the revival peak times  $t_{r}$ satisfy a simple relation 
\begin{equation}
(\Omega_{n+1}-\Omega_{n})t_{r}=2\pi m \hspace{2em}(m=1,2,3,\cdots), \label{t-cr}
\end{equation}
namely,  the neighbouring oscillation terms differ by an integer times  $2\pi$. 
For a large bath size, i.e.  $N \gg1$, we get the location of the $m$-th revival peak in time 
%\begin{equation}
$t_{r} \simeq\frac{\pi N}{A} m$,
%\end{equation}
which is confirmed in Fig.~\ref{Fig2:cr-entropy}.  It is linearly proportional to the bath spin number $N$. 
In Fig.~\ref{Fig4:cr}, we will discuss how the initial revival dynamics remains almost unaffected by a relatively large  inhomogeneity, $\alpha \sim 1$. This CR dynamics is distinct from the revivals observed in the spin-echo signal of central spin systems, when the evolution time is a multiple of the nuclear Larmor periods~\cite{Childress:2006,Bluhm:2011,Cywinski:2009} (thus, that revival time is independent of $N$). 

{\bf Decoherence and  entanglement.} In order to characterize  the nature of entanglement  between the bath and the central spin, we further calculate the  reduced density matrix of the central spin by tracing out the degrees of  freedom of the bath spins $ \left\lbrace \vert n \rangle \langle n \vert \right\rbrace$ 
\begin{equation}
 \rho_{cs}=\Tr_{ \left\lbrace \vert n \rangle \langle n \vert \right\rbrace }\left[ \vert \psi(t) \rangle  \langle \psi(t) \vert \right] 
 =\left(
\begin{array}{cc}
A(t) & B(t) \\
B(t) ^{*}& 1-A(t)\\
\end{array}
\right),
\end{equation}
where the matrix elements  read
\begin{eqnarray}
A(t)&=&\sum_{n=0}^{N}{ C_{N}^{n}}[\sin^{2}(\theta/2)]^{n}[\cos^{2}(\theta/2)]^{N-n} \vert  P_{\uparrow}^{n} \vert^2, \nonumber \\
B(t)&=&\sum_{n=0}^{N}\sqrt{ C_{N}^{n+1}C_{N}^{n}} [\sin^{2}(\theta/2)]^{n+\frac{1}{2}}[\cos^{2}(\theta/2)]^{N-n-\frac{1}{2}} \nonumber \\
&& \times P_{\uparrow}^{n+1} (P_{\downarrow}^{n})^{*} \nonumber .
%C(t)&=&B(t) ^{*}. 
\end{eqnarray}
%Here the parameters are defined as the above, see SM \cite{SuppM}.
%[Stefano] slight rephrasing
The  purity and Von Neumann entropy,  which characterize the entanglement between the central spin and the bath spins, are given explicitly via the relations $ \gamma\equiv \Tr[\rho_{cs}^2] $ and $S(\rho_{cs})\equiv-\Tr[\rho_{cs} \ln\rho_{cs}]$. They are displayed in Fig.~\ref{Fig2:cr-entropy}(c,d), showing an important decoherence effect: while the central spin entropy (purity) is initially small (large) at the CR points, it gradually increases (decreases) with time. As a consequence, the CR gradually vanish in the long time limit.
 %

%\begin{widetext}
\begin{figure}
  \begin{center}
%\begin{flushleft}
\includegraphics[scale=0.075]{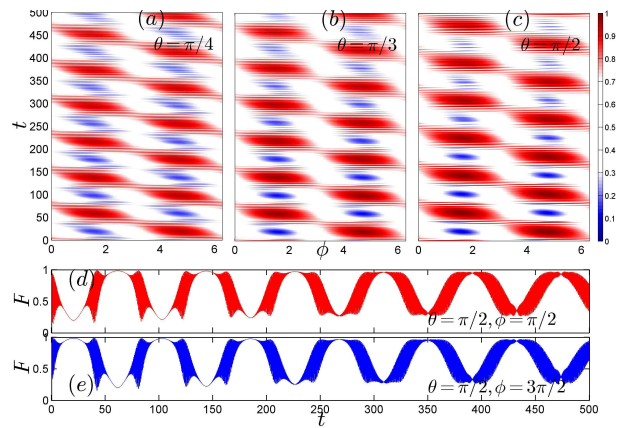}
%\includegraphics[scale=0.35]{Qcrfig4.eps}
%\end{flushleft}
\end{center} 
\caption{ Contour  plots of fidelity vs the phase $\phi$ for different values of $\theta$.  (a)-(c):  time evolution of the fidelity of  the  reduced density matrix of  the  central spin $\rho_{cs}$ against the state $\vert \phi \rangle=\frac{1}{\sqrt{2}}[\vert \uparrow \rangle_{0}+e^{-i\phi}\vert \downarrow \rangle_{0}]$.  Panels (d) and (e)  show time evolution of the the fidelity for the phase angles $\phi=\pi/2$ and $\phi=3\pi/2$. In all panels, $B=\Delta=A=1$.}
\label{Fig3-fidelity}
\end{figure} 

%[Stefano] shortened text
We further  look at the  fidelity of $\rho_{cs}$ with  respect to  the state $\vert\phi \rangle=\frac{1}{\sqrt{2}}(\vert \uparrow \rangle_{0}+e^{-i\phi}\vert \downarrow \rangle_{0})$.  By definition,  the fidelity  is  given  by 
  $F(\rho_{cs},\rho_{\phi})=\sqrt{\langle \phi \vert \rho_{cs} \vert\phi \rangle}$ for the pure state  $\rho_{\phi}$,
leading to 
\begin{equation}
F=\sqrt{ [1+Be^{-i\phi}+B^{*}e^{i\phi}]/2}.\label{Fidelity}
\end{equation}
In Fig.~\ref{Fig3-fidelity}, we contour plot $F$ in  the phase-time plane $(\phi,t)$. 
 It is interesting to observe that  fidelity oscillations with high contrast occur for the two special values of the phase $\phi=\frac{\pi}{2}$ or $\phi=\frac{3\pi}{2}$. 
As seen from Fig.~\ref{Fig3-fidelity}(a-c), the optimal choices of $\phi$ are independent of $\theta$ \cite{SuppM}. The fidelity peaks occur around the middle points of the collapse regions and are over $92\%$ in Fig.~\ref{Fig3-fidelity}(d-e), where $\theta=\pi/2$. Instead, moving away from $\theta=\pi/2$ causes a reduction of the maximum fidelity. The  slow decay of the oscillations means that the central spin is able to decouple periodically from the bath. A longer decoherence time  may  facilitate dynamical control of entangled states in NV center devices \cite{RBLiu:2017}.

%
%In addition, the coherence factor of the XXZ central spin problems   is defined by 
%\begin{equation}
%S^{-}_{0}(t)=\langle \psi(t) \vert \mathbf{\bf s}^{-}_{0}\vert \psi(t) \rangle=\Tr[ \rho_{cs} \mathbf{\bf s}^{-}_{0}], 
%\end{equation}
% which  also describes  the coherence behaviour of the central spin. 
%The  coherence factor can be derived from the reduced density matrix $\rho_{cs}$ too. 
%
%The square norm of  the coherence factor is given by 
%\begin{equation}
%$\vert S^{-}_{0}(t) \vert^{2}=\vert B(t)\vert^{2}$ for the model (\ref{Ham-2}), see  the SM \cite{SuppM} for a detailed discussion. 
%\end{equation}

%\end{widetext}

%\begin{figure}[ht]
  %\begin{center}
%%\begin{flushleft}
%\includegraphics[scale=0.35]{Qcrfig4.eps}
%\end{flushleft}
%\end{center} 
%\caption{First three subplots, two times spin polarization of central spin evolve with time at different bath size $N=100,500,1000$, detuning term $\Delta_{n}=0$(namely $B=\Delta=0$), transverse coupling $A=1/\sqrt{N}$, initial state angle satisfying $N\sin^{2}(\theta/2)=25$. The fouth subplot, the inversion of the Janynes-Cummings model, detuning $\Delta_{JC}=0$, average photons number $ \langle n \rangle=25$, seeing the M.O.Scully et al's book, \emph{quantum optics}.}
%\label{Qcrfig4}
%\end{figure} 

{\bf Statistical nature of the  generalized Jaynes-Cummings model.} We now give an exact  mapping between the homogenous XXZ  central spin problems  and the atom-field interaction  model in quantum nonlinear optics. 
From the  Holstein-Primakoff transformation $\mathbf{J}^{+}=\sqrt{N}\mathbf{\bf a}^{\dag}\sqrt{1-\mathbf{\bf a}^{\dag}\mathbf{\bf a}/N}$, 
$ \mathbf{J}^{-}=\sqrt{N}\sqrt{1-\mathbf{\bf a}^{\dag}\mathbf{\bf a}/N}\mathbf{\bf a}$, and $ \mathbf{J}^{z}=-\frac{N}{2}+\mathbf{\bf a}^{\dag}\mathbf{\bf a}$, where $\mathbf{\bf a}$ ($\mathbf{\bf a}^{\dag}$) is a bosonic annihilation (creation) operator, 
we may  build a deep  connection  between the  central spin problems and the matter-light interaction systems \cite{Scully:1997}.  
In the  large $N$ limit,   the Hamiltonian Eq. (\ref{Ham-2}) becomes (up to a constant)
\begin{equation}
H=B^{\prime}\mathbf{\bf s} ^{z}_{0}+\sqrt{N} A\left[ \mathbf{\bf s}_{0}^{+}\mathbf{a}+ \mathbf{\bf s}_{0}^{-} \mathbf{\bf a}^{\dag}\right]+2\Delta\mathbf{\bf s}_{0}^{z} \mathbf{\bf a}^{\dag}\mathbf{\bf a}+h\mathbf{\bf a}^{\dag}\mathbf{\bf a}, \label{Ham-3}
\end{equation}
 where the effective magnetic field  is  $B^{\prime}=B+h$ and $h$ is related  to  the light frequency \cite{note}, 
whereas $\sqrt{N} A $ is related to the coupling constant between the atom and bosonic mode \cite{Scully:1997,Bogoliubov,Amico:2005}.
This model (\ref{Ham-3})  can be regarded as a generalized Jaynes-Cummings (JC)  model,  in which the atomic transition frequency  also depends on the number of photons. 
For $\Delta=0$, the  central spin problem Hamiltonian  (\ref{Ham-2}) exactly reduces to  the JC model.

In order to see a dynamical connection between the two systems, let's define the  inversion $W_{cs}(t)$  of the central spin following  the JC model. The inversion is immediately found from Eq.~(\ref{eqnarray}), since $W_{cs}(t)=\langle \psi(t)\vert \boldsymbol{\sigma}_{z}\vert \psi(t) \rangle=2S_{0}^{z}$.
Using the  Poisson limit theorem  when $N\rightarrow \infty$ and $p\rightarrow 0$,  we have 
%\begin{equation}
$C_{N}^{n} p^{n}(1-p)^{N-n} \simeq e^{- \lambda} \frac{\lambda^{n}}{n!}$,
%\end{equation}
where  $\lambda=Np$.
Then,  taking   the  limit $\theta \rightarrow 0 $ for the initial angle,   we find $C_{N}^{n}[\sin^{2}(\theta/2)]^{n}[\cos^{2}(\theta/2)]^{N-n}   \simeq   e^{- \zeta^{2}} \frac{ (\zeta^{2})^n}{n!}$, where  $\zeta^{2}=N\sin^{2}(\theta/2)$.
 As a consequence,  the  inversion  is given by 
\begin{eqnarray}
&W_{cs}(t)&=\sum_{n=0}^{\infty} e^{- \zeta^{2}} \frac{ (\zeta^{2})^n}{n!}   \nonumber \\
&&   \times \left[\frac{ \Delta^{2}_{n+1}}{(\Omega_{n+1})^2}+\frac{4(n+1)NA^2}{(\Omega_{n+1})^2} \cos(\Omega_{n+1}t) \right]. ~~~~
\label{JC-inversion}
\end{eqnarray}
Here the parameters read
$ \Delta_{n+1}=B-N\Delta$ and $ \Omega_{n+1}=\sqrt{\Delta_{n+1}^2+4(n+1)NA^2}$.
This is nothing but an exact result of quantum CR of  the generalized JC   model (\ref{Ham-3}).
Moreover, taking the limit  $\Delta \rightarrow 0$, the expression Eq.~(\ref{JC-inversion}) recovers Eq.~($6.2.21$) of Ref.~\cite{Scully:1997} for  the JC  model,  see also  Fig.~S10  in the SM \cite{SuppM}. The exact correspondence between the special case of the XXZ central spin problems (\ref{Ham-2})  and  the  JC    model reads:
\begin{equation}
\begin{array}{ccc}
Central \hspace{0.2em} spin \hspace{0.2em} model & ~~& JC \hspace{0.2em} model \\
B  & \rightarrow &  \Delta_{JC} \\
\sqrt{N}A &\rightarrow  & g \\
\zeta^{2} &\rightarrow  & \langle n \rangle 
\end{array}
\end{equation}
where the parameters of the Jaynes-Cummings  model  are respectively the detuning, the coupling between the photon and atom,  and  the average  photon  number. This correspondence presents a deep relation  between the two types of models, i.e.,  a  large number of  bath spins with   a particular choice of the spin coherent state ($\theta \to 0$)  can be regarded as a single-occupied-multilevel fermionic  field  that naturally reduces to 
a bosonic  field, revealing a statistical nature of Holstein-Primakoff transformation.

\begin{figure}
\begin{center}
%\begin{flushleft}
\includegraphics[width=0.475\textwidth]{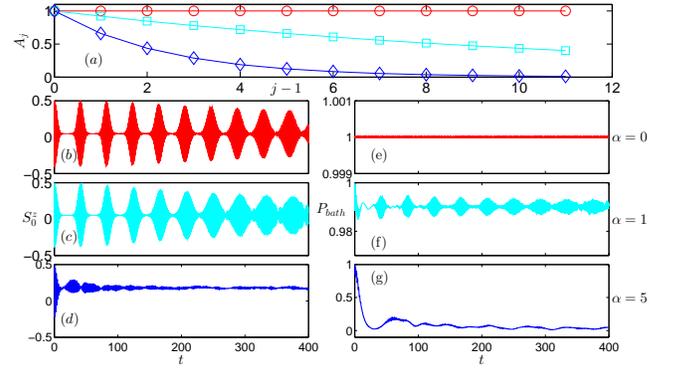}
%\end{flushleft}
\end{center} 
\caption{Effect of inhomogeneity on the CR dynamics. (a): Coupling strengths $A_{j}=A \exp(-\alpha(j-1)/N)$, with $\alpha =0, \, 1,\, 5$, corresponding to homogeneous, intermediate, and strongly inhomogeneous couplings, respectively.  (b)-(d): Time evolution of  $S_0^z$  obtained from Eq.~(\ref{Ham-1})  using $\Delta_j=A_j,\, A=1,\, N=12$,  $B=1$, and the three chosen values of $\alpha$. (e)-(g): Expectation values  of the collective bath projector.}
\label{Fig4:cr}
\end{figure} 

{\bf Inhomogeneous central spin problem.}  In order to assess the effect of inhomogeneity on the quantum dynamics discussed so far, we  diagonalize Eq.~(\ref{Ham-1}) with different values of the inhomogeneity parameter $\alpha$. The wave function and central spin polarization for the inhomogeneous model, at arbitrary $M$ and time, are obtained in the SM \cite{SuppM}. We  observe that a strong inhomogeneity of the coupling amplitudes leads to a breakdown of  CR dynamics, see  Fig.~\ref{Fig4:cr}(d). However, we also find that with $\Delta_j=A_j$ the revivals are remarkably robust to moderate values of the inhomogeneity, $\alpha \sim 1$. 

To shed some light to the origin of such behavior, we introduce the collective bath projector $\hat{P}_{bath}=\sum_{n} \vert n \rangle \langle n \vert$ onto the states with maximal eigenvalue of ${\bf J}^2$. For a large inhomogeneity factor $\alpha$, the expectation value
\begin{equation}
P_{bath}(t)= \langle  \hat{P}_{bath} \rangle =\sum_{n} \vert \langle n \vert \psi(t) \rangle \vert^{2}
\label{collective}
 \end{equation}
quickly decays to small values, see Fig.~\ref{Fig4:cr}(g). Instead, persistence of $P_{bath} \sim 0.99$ shown in Fig.~\ref{Fig4:cr}(f) corresponds to the robust CR dynamics of panel (c). Interestingly, the decay of $P_{bath}(t)$ with inhomogeneous couplings depends sensitively on the isotropy of the interactions, being much quicker in the Ising case  ($A_j=0$) \cite{SuppM}.  We also obtain evidence from the numerical simulations ($N \leq 12 $) that the CR dynamics at given $\alpha \neq 0$ persists for a longer time with larger number of bath spins. These observations reveal the subtle dependence of CR with respect to the Hamiltonian  parameters. In the SM \cite{SuppM} we also explore how CR signatures appear by gradually increasing the degree of bath spin polarization.

%\begin{widetext}
%\begin{figure}
%  \begin{center}
%%\begin{flushleft}
%\includegraphics[width=0.49\textwidth]{Fig1-revial-alpha-m.eps}
%%\end{flushleft}
%\end{center} 
%\caption{\color{blue} The  central spin polarization  $S_0^z$ evolves in time  under  different  values of the inhomogeneous  couplings: (a) shows the coupling strengths of  $A_{j}=A \exp(-\alpha(j-1)/N)$ with $\alpha =0, \, 1.0,\, 5.0$, corresponding to the  homogeneous, weakly and strongly inhomogeneous couplings, respectively.  (b)-(d) show  the corresponding  time evolutions of  $S_0^z$ for these values, which  are obtained from  diagonalization of the Hamiltonian (\ref{Ham-1})  with the setting    $\Delta_j=A_j,\, A=1,\, N=12$ and magnetic field $B=1$. }
%\label{Fig1:cr}
%\end{figure} 
%%\end{widetext}
%	
%\begin{figure}
%  \begin{center}
%%\begin{flushleft}
%\includegraphics[width=0.475\textwidth]{Fig5_J2proj.eps}
%%\end{flushleft}
%\end{center} 
%\caption{\color{blue} The bath spin  angular momentum ratio(left column) and the expectation of collective bath projector (right column) with inhomogeneity factor $\alpha =0,\, 1.0,\,3.0,\, 5.0$. The Hamiltonian (\ref{Ham-1})  with the setting    $\Delta_j=A_j,\, A=1,\, N=12$ and magnetic field $B=1$. }
%\label{Fig5:JP}
%\end{figure} 

In summary we have obtained  the benchmark quantum dynamics of the XXZ central spin problem  with homogenous and inhomogeneous coupling amplitudes.
Analytical results of quantum CR, entanglement entropy and  fidelity  provide  rich insights into quantum dynamic control of entangled states  for quantum metrology. 
The effects of inhomogeneity on the robustness and decay of  the CR have been studied as well.  
   Our   methods can be directly applied to high central spin problems as well as models of multiple atoms coupled to a cavity in quantum nonlinear optics. 
%

%---------------------------------------Figure-----------------------------------------------

%---------------------------------------Figure-----------------------------------------------

\noindent

{\em Acknowledgments.} We thank  Henrik Johannesson, Chao-Hong Lee, Shi-Zhong Zhang and Li You  for helpful discussions.
This work is supported by   the National Key R\&D Program of China  No. 2017YFA0304500,  the key NSFC grant No.\ 11534014 and NSFC grant No. 11874393.
 SC acknowledges support from the National Key R\&D Program of China No. 2016YFA0301200 and NSFC (Grants No. 11574025, No. 11750110428 and No. 1171101295).
HQL and SC acknowledge financial support from NSAF U1530401 and computational resources from the Beijing Computational Science Research Center (CSRC). XWG thank CSRC for kind hospitality, where part of this paper was prepared.

% end here if supplement is not included
\ifNOSUP\end{document}\else%

%%%%%%%%%%%%%%%%%%%%%%%%%%%%%%%%%%%%%%%%%%%%%%%%%%%%%%
%
% Supporting Material
%
%%%%%%%%%%%%%%%%%%%%%%%%%%%%%%%%%%%%%%%%%%%%%%%%%%%%%%
\clearpage\newpage
\setcounter{figure}{0}
\setcounter{table}{0}
\setcounter{equation}{0}
\def\thefigure{S\arabic{figure}}
\def\thetable{S\arabic{table}}
\def\theequation{S\arabic{equation}}
\setcounter{page}{1}
\pagestyle{plain}
%%%%%%%%%%%%%%%%%%%%%%%%%%%%%%%%%%%%%%%%%%%%%%%%%%%%%%
\begin{widetext}

\section*{Supporting Material for ``Exact  quantum dynamics    of XXZ  central spin problems''}

\begin{center}
\noindent{Wen-Bin He, Stefano Chesi, H.-Q. Lin, Xi-Wen Guan}
\end{center}

\maketitle
The evolution of wave function is the first step  to access quantum dynamics of the considered many-body system.
 In this supplementary material, we present in detail the derivation of the wave function of homogeneous XXZ central spin problems. 
Using the obtained exact wave function, we further derive the time evolution of  important physical quantities, like the spin polarization, quantum purity, Von Neumann entropy, coherence factor of the central spin, etc.
 These results not only provide  benchmark dynamics of the XXZ central spins problems but also build  an  exact dynamical connection  with models in quantum nonlinear optics, such as  the Jaynes-Cummings model, etc. 
 The methods developed here can be extended to other cases, for example, high central spin problems, multiple atoms coupled to a bosonic mode in quantum nonlinear optics, etc. Finally, we discuss the influence of inhomogeneous coupling to the collapse and revival. 

\section{Time evolution of the  wave function}
% \subsection{The wave function }
The Hamiltonian of the homogeneous XXZ central spin model can be written as 
 \begin{equation}
 H=B\mathbf{\bf s} ^{z}_{0}+2\sum_{j=1}^{N}[A(\mathbf{\bf s}_{0}^{x} \mathbf{\bf s}_{j}^{x}+\mathbf{\bf s}_{0}^{y} \mathbf{\bf s}_{j}^{y})+\Delta\mathbf{\bf s}_{0}^{z} \mathbf{\bf s}_{j}^{z} ].
 \end{equation}
 For our convenience, in the following derivation we introduce the large spin operator $\mathbf{J}=\sum_{j=1}^{N}\mathbf{s}_{j}$. Then, the  Hamiltonian is transformed into the form
\begin{equation}
H=B\mathbf{\bf s} ^{z}_{0}+A(\mathbf{\bf s}_{0}^{+} \mathbf{\bf J}^{-}+\mathbf{\bf s}_{0}^{-} \mathbf{\bf J}^{+})+2\Delta\mathbf{\bf s}_{0}^{z} \mathbf{\bf J}^{z}.
\label{Hhcs}
\end{equation}
 If the bath spins are prepared in  a  spin coherent state $ \vert \theta \rangle=\otimes_{j=1}^{N}[\sin(\theta/2)\vert \uparrow \rangle_{j}+\cos(\theta/2) \vert \downarrow \rangle_{j}] $, which can be  written in terms of Dicke states  as $\vert \theta \rangle= \sum_{n=0}^{N}\sqrt{C_{N}^{n}}[\sin(\theta/2)]^{n}[\cos(\theta/2)]^{N-n}\vert n \rangle$ ( $\vert n \rangle= \vert \frac{N}{2},n- \frac{N}{2}\rangle$ is the eigenstate of $\mathbf{J}^{2}$ and $\mathbf{J}^{z}$), the initial state reads
\begin{equation}
\vert  \Phi_{0}\rangle=\vert \uparrow \rangle_{0}\otimes \left[\sum_{n=0}^{N}\sqrt{C_{N}^{n}}[\sin(\theta/2)]^{n}[\cos(\theta/2)]^{N-n}\vert n \rangle \right].
\label{initial}
\end{equation}
We first assume, like in (\ref{initial}), that the central spin is in the up state, in order to derive the mapping between the central spin problem (\ref{Hhcs}) and  the Jaynes-Cummings model given by M. O. Scully et al., \emph{Quantum Optics} \cite{Scully:1997}. If the central spin is initially in the the down state, the solution can be derived in a similar manner and is presented in Eq.~(\ref{psi_down}). 

Due to the unitary evolution of wave function, we have 
\begin{align}
&\vert \psi(t) \rangle= e^{-iHt}\vert \Phi_{0} \rangle=\sum_{m=0}^{\infty}(-it)^m H^m/{m!}  \vert \Phi_{0} \rangle  \nonumber \\
& =\sum_{n=0}^{N}\sqrt{ C_{N}^{n}} [\sin(\theta/2)]^{n}[\cos(\theta/2)]^{N-n} \sum_{m=0}^{\infty}(-it)^m H^m/{m!} \vert \uparrow \rangle_{0} \vert n \rangle.
\label{Qwfep}
\end{align}
Using the following eigenstate relation of angular momentum operators $\mathbf{J}^{2}, \mathbf{J}^{z}$ 
\begin{align*}
&\mathbf{J^{-}}  \vert n \rangle=  \sqrt{b_{n}} \vert n-1 \rangle,\\
& \mathbf{J^{+}} \vert n \rangle= \sqrt{b_{n+1}} \vert n+1 \rangle,
\end{align*}
where $b_{n}=n(N-n+1)$,  we obtain: 
\begin{align*}
&H \vert \uparrow \rangle_{0} \vert n \rangle= w_{n} \vert \uparrow \rangle_{0} \vert n \rangle +\sqrt{b_{n+1}}  \vert \downarrow \rangle_{0}\vert n+1 \rangle, \\
&H \vert \downarrow \rangle_{0}\vert n \rangle=  -w_{n} \vert \downarrow \rangle_{0}\vert n \rangle +\sqrt{b_{n}}\vert \uparrow \rangle_{0} \vert n-1 \rangle,
\end{align*}
where we denoted $w_{n}=B/2+(n-N/2)\Delta$.  By defining $S_{m}=H^m \vert \uparrow \rangle_{0} \vert n \rangle$ and $t_{m}=H^m \vert \downarrow \rangle_{0} \vert n \rangle$, we further derive the following recurrence relations by  applying  the Hamiltonian  $m$ times on the Dicke states $\vert \uparrow \rangle_{0} \vert n \rangle,\vert \downarrow \rangle_{0}\vert n \rangle $
\begin{align}
& S_{m+2}+(w_{n+1}-w_{n})S_{m+1}-(b_{n+1}A^2+w_{n}w_{n+1})S_{m}=0, \\
& t_{m+2}+(w_{n}-w_{n-1})t_{m+1}-(b_{n}A^2+w_{n-1}w_{n})t_{m}=0. 
\end{align}
Here the calculation is rather involved but straight forward. 
The initial conditions for the above recurrence relations read 
\begin{align*}
 S_{0}=\vert \uparrow \rangle_{0} \vert n \rangle, \hspace{0.5em} S_{1}=w_{n}\vert \uparrow \rangle_{0} \vert n \rangle+\sqrt{b_{n+1}}A \vert \downarrow \rangle_{0} \vert n+1 \rangle, \\
  t_{0}=\vert \downarrow \rangle_{0} \vert n \rangle, \hspace{0.5em} t_{1}=-w_{n}\vert \downarrow \rangle_{0} \vert n \rangle+\sqrt{b_{n}}A \vert \downarrow \rangle_{0} \vert n-1 \rangle. \\
\end{align*}
In view of the  characteristic equation of recurrence relation  $S_{m}$
\begin{equation}
\lambda^2+(w_{n+1}-w_{n})\lambda-(b_{n+1}A^2+w_{n}w_{n+1})=0,
\end{equation}
we obtain the characteristic roots of above equation
\begin{align*}
\lambda_{1,2}(n)=\frac{(w_{n}-w_{n+1}) \pm \sqrt{(w_{n}+w_{n+1})^2+4b_{n+1}A^2}}{2}.
\end{align*}
Using the above initial conditions,  the series  $S_{m}$ is given by 
\begin{equation}
H^{m}\vert \uparrow \rangle_{0} \vert n \rangle=S_{m}=\frac{S_{1}-\lambda_{2}(n)S_{0}}{\lambda_{1}(n)-\lambda_{2}(n)}\lambda_{1}^{m}(n)+\frac{\lambda_{1}(n)S_{0}-S_{1}}{\lambda_{1}(n)-\lambda_{2}(n)}\lambda_{2}^{m}(n).
\end{equation}

We further  obtain series $t_{m}$ by the same method. We only need replacing the initial conditions  $ S_{0} $ and $S_{1}$ with $  t_{0}$ and $t_{1}$, leading to: 
\begin{equation}
H^{m}\vert \downarrow \rangle_{0} \vert n \rangle =\frac{t_{1}-\lambda_{2}(n-1)t_{0}}{\lambda_{1}(n-1)-\lambda_{2}(n-1)}\lambda_{1}^{m}(n-1)+\frac{\lambda_{1}(n-1)t_{0}-t_{1}}{\lambda_{1}(n-1)-\lambda_{2}(n-1)}\lambda_{2}^{m}(n-1). 
\end{equation}
The functions  $S_{m}$ and $t_{m}$ are actually related in the following way:
\begin{equation*}
t_{m}(n+1)=S_{m}(n).
\end{equation*} 

By substituting the expression of  $S_{m}$ into Eq.~(\ref{Qwfep}), we obtain the wave function 
\begin{align}
\vert \psi(t) \rangle=\sum_{n=0}^{N}\sqrt{ C_{N}^{n}}  [\sin(\theta/2)]^{n}[\cos(\theta/2)]^{N-n} \left[  \frac{S_{1}-\lambda_{2}S_{0}}{\Omega_{n+1}}\exp[-i\lambda_{1}t]+\frac{\lambda_{1}S_{0}-S_{1}}{\Omega_{n+1}}\exp[-i\lambda_{2}t]  \right].
\end{align}
Substituting the initial condition $ S_{0},S_{1}$ and the characteristic roots into  the above formula, we obtain  the wave function at  arbitrary times 
\begin{equation}
|\psi(t)\rangle =e^{-i\theta(t)}\cdot \sum_{n=0}^{N}\sqrt{C_{N}^{n}}  [\sin(\theta/2)]^{n}[\cos(\theta/2)]^{N-n} \left[ P_{\downarrow}^{n} (t) \vert \downarrow \rangle_{0} \vert n+1\rangle+P_{\uparrow}^{n}(t) \vert \uparrow \rangle_{0} \vert n\rangle\right]. \label{Qcrwf}
\end{equation}
Here the  global phase $\theta(t)$ can be omitted and the two  amplitudes  are given by 
\begin{align*}
&P_{\uparrow}^{n}= -i\frac{\Delta_{n+1}}{ \Omega_{n+1}}\sin( \frac{\Omega_{n+1}t}{2})+\cos( \frac{\Omega_{n+1}t}{2}), \\
& P_{\downarrow}^{n}=-i\frac{2\sqrt{b_{n+1}}A}{ \Omega_{n+1}}\sin( \frac{\Omega_{n+1}t}{2}).
\end{align*}
In the above equations, the parameters were denoted by 
\begin{align}
&b_n=n(N-n+1), \nonumber \\
&\Delta_{n}=B+(2n-1-N)\Delta, \nonumber \\
& \Omega_{n}=\sqrt{\Delta_{n}^2+4b_nA^2}.
\label{Qsmpa}
\end{align}
It is worth noting that the roots $\lambda_{1,2}$ are the eigenvalues  of the Hamiltonian (\ref{Hhcs}). After obtaining  the  wave function, we can derive spin polarization, reduced density matrix, quantum purity, Von Neumann entropy, etc. We will discuss  these properties below. 

%%%%%%%%%%%%%

\begin{figure}
\begin{center}
\includegraphics[scale=0.4]{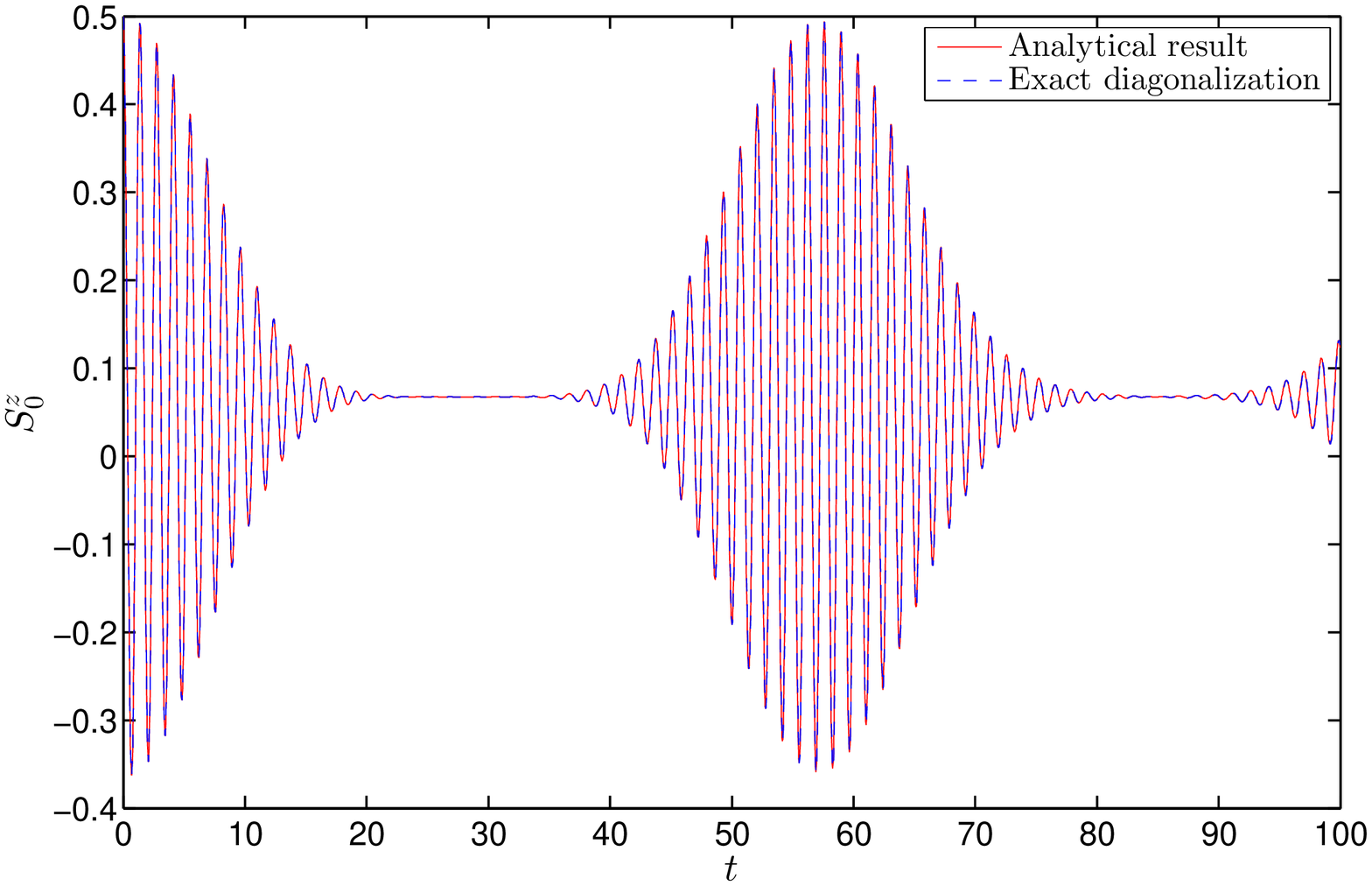}
\end{center} 
\caption{Comparison between our analytical result of the central spin polarization,  Eq.(\ref{Qcrsz0})  (solid red line), and the exact diagonalization result (dashed blue line). Here we set the bath spins  $N=8$, magnetic field $B=0.5$, coupling $A=\Delta= 0.5$.}
\label{Qcrsmfig1}
\end{figure}

\section{ Main  results}

\emph{Spin polarization}.

By using above wave function,  Eq.~(\ref{Qcrwf}), we  can get spin polarization and reduced density matrix of the  central spin, which are defined as below
\begin{align}
& S^{z}_{0}(t)= \langle \psi(t) \vert  \mathbf{s}^{z}_{0}\vert \psi(t) \rangle, \\
& \rho_{cs}=\Tr_{ \left\lbrace \vert n \rangle \langle n \vert \right\rbrace }\left[ \vert \psi(t) \rangle  \langle \psi(t) \vert  \right].
\label{Qcrsmrho}
\end{align}
We concentrate on the spin polarization of central spin $S^{z}_{0}(t)$. The  interesting result is that spin polarization of the  central spin displays quantum collapse and  revivals,  like  the inversion of  the  Jaynes-Cummings model \cite{Scully:1997}.
 Although  the phenomenon of quantum collapse and revival  had been numerically studied in  Ref.~\cite{QcrPra}, here  we obtain  the exact  form of the  quantum collapse and revival in the homogeneous XXZ central spin model,  given by the following expression
\begin{align}
& S_{0}^{z}(t)=\frac{1}{2}\sum_{n=0}^{N}C_{N}^{n}[\sin^{2}(\theta/2)]^{n}[\cos^{2}(\theta/2)]^{N-n}  \left[ \frac{ \Delta^{2}_{n+1}}{(\Omega_{n+1})^2}+\frac{4b_{n+1}A^2}{(\Omega_{n+1})^2} \cos(\Omega_{n+1}t) \right],
\label{Qcrsz0}
\end{align}
where  parameters are defined as  in the  previous Eq.~(\ref{Qsmpa}). A comparison between our analytical result Eq.~(\ref{Qcrsz0}) and exact diagonalization  shows a perfect agreement, see Fig.~\ref{Qcrsmfig1}.   Here we observe that  the longitudinal interaction $\Delta$  facilitates  the quantum  revival even for small bath size.  For  example, such  revivals can be observed for  $N=4, \,6,\,8 $, see Fig.~\ref{Qcrsmfig2}. 
This observation could help  the  experimental realization of  the  quantum collapse and revival through quantum devices, such as superconducting circuits~\cite{nature}.

\begin{figure}
\begin{center}
\includegraphics[scale=0.4]{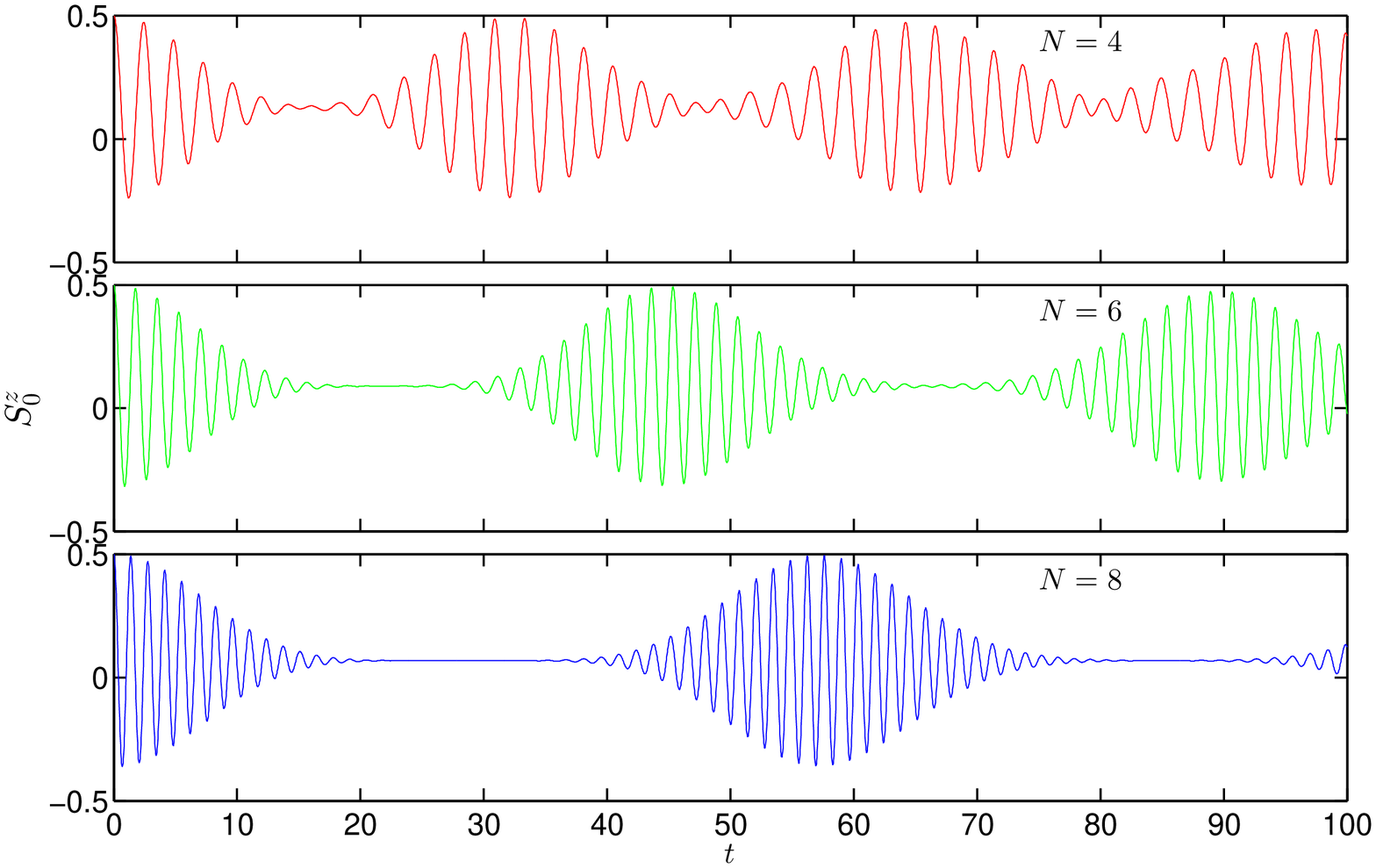}
\end{center} 
\caption{ Spin  polarization of  the  central spin for small bath size $N=4,6,8$. The   magnetic field and coupling are  $B=\Delta=A=0.5$.}
\label{Qcrsmfig2}
\end{figure}

\begin{figure}[h!]
\begin{center}
\includegraphics[scale=0.35]{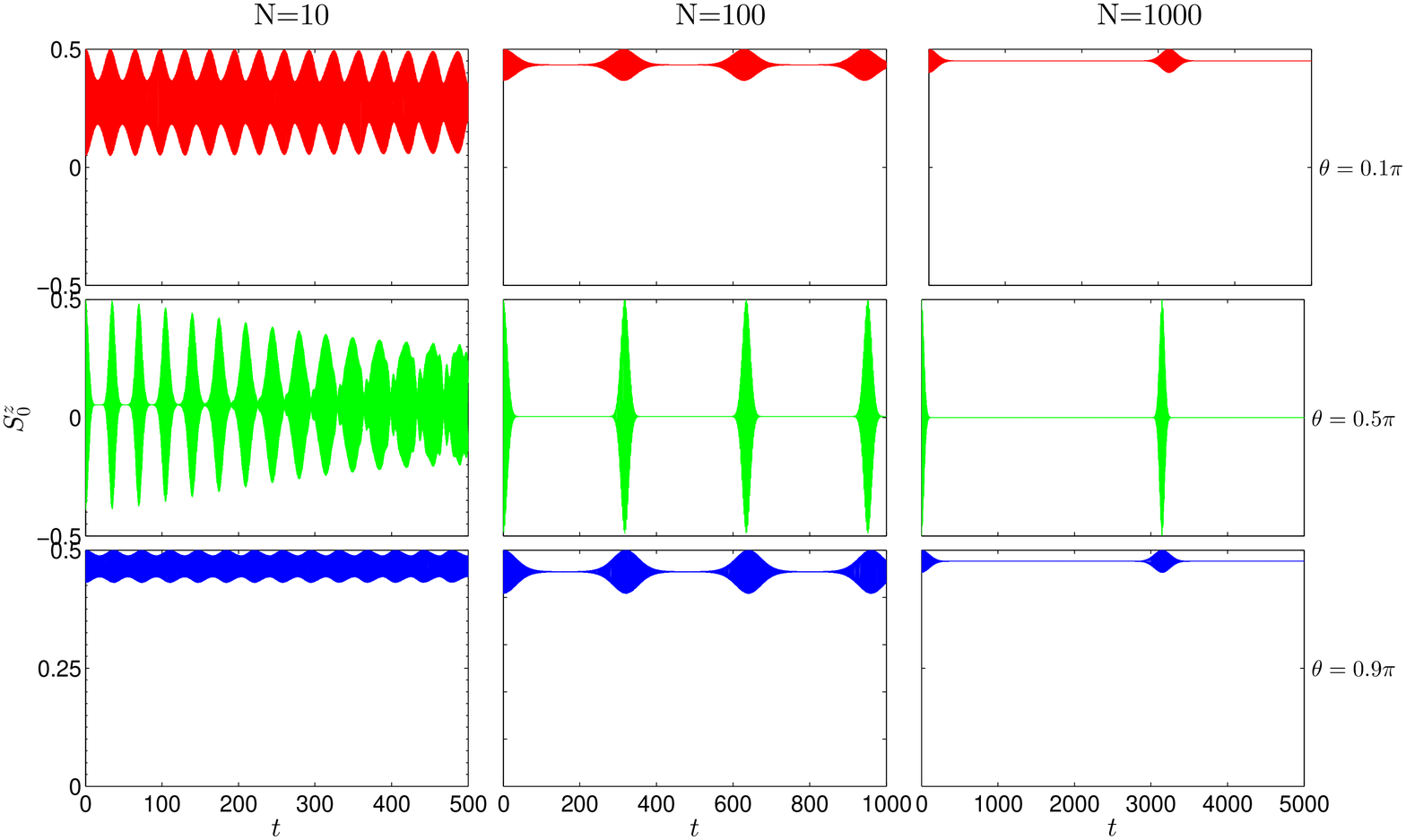}
\end{center} 
\caption{Spin  polarization of the  central spin for different bath  sizes  and initial state angles.   Magnetic field and coupling  are   $B=\Delta=A=1$.}
\label{Qcrsmfig3}
\end{figure} 

As discussed in the main text, the resonant condition $B=\Delta=A$ is most favorable to observe the collapse and revival dynamics. However,  the phenomenon is also significantly affected by the number of bath spins $N$ and the initial value of the polarization. We show some examples in Fig.~\ref{Qcrsmfig3}.

If the central spin is prepared in the spin-down state, namely the initial state is  $\vert  \Phi_{0}\rangle=\vert \downarrow \rangle_0 \otimes \vert  \theta\rangle$, the wave function at arbitrary  times  is given by 
 \begin{align}\label{psi_down}
\vert \psi(t) \rangle  = &\sum_{n=0}^{N}\sqrt{C_{N}^{n}}  [\sin(\theta/2)]^{n}[\cos(\theta/2)]^{N-n} \left[ \left( i\frac{\Delta_{n}}{ \Omega_{n}}\sin( \frac{\Omega_{n}t}{2})+\cos( \frac{\Omega_{n}t}{2}) \right) \vert \downarrow \rangle_{0} \vert n\rangle\right. \\ \nonumber 
& \left.  -i \frac{2\sqrt{b_{n}}A}{ \Omega_{n}}\sin( \frac{\Omega_{n}t}{2})  \vert \uparrow \rangle_{0} \vert n-1\rangle\right].
\end{align}
Thus the spin polarization of central spin obeys the  evolution
 \begin{equation}
S_{0}^{z}(t)=-\frac{1}{2}\sum_{n=0}^{N}C_{N}^{n}[\sin^{2}(\theta/2)]^{n}[\cos^{2}(\theta/2)]^{N-n}\  \left[ \frac{ \Delta^{2}_{n}}{(\Omega_{n})^2}+\frac{4b_{n}A^2}{(\Omega_{n})^2} \cos(\Omega_{n}t) \right].
\label{Qcrsz0down}
\end{equation}

\bigskip
 \noindent \emph{Effect of Randomness in Bath spins}.

The density matrix of the spin coherent state is chosen as   $\rho(\theta,\varphi)=\vert \theta,\varphi \rangle  \langle \theta,\varphi \vert $, here  
  the spin coherent state is given by 
\begin{eqnarray} 
&\vert \theta,\varphi \rangle=& \otimes_{j=1}^{N}[\sin(\theta/2)e^{-i \varphi}\vert \uparrow \rangle_{j}+\cos(\theta/2)\vert \downarrow  \rangle_{j}] \nonumber \\
&=& \sum_{n=0}^{N}\sqrt{C_{N}^{n}}[\sin(\theta/2)]^{n}e^{-i n \varphi} [\cos(\theta/2)]^{N-n} \vert n \rangle,
\end{eqnarray}
here $ \vert n \rangle$ denotes Dicke state. 
We take the ensemble average of the density matrix $\rho(\theta,\varphi)$ over  the solid angles
\begin{equation}
\rho^{\prime}=\int \rho(\theta,\varphi) \frac{d\Omega}{4 \pi}=\int \rho(\theta,\varphi) \frac{\sin(\theta)d\theta d\varphi} {4 \pi}
\end{equation}
We thus  get 
\begin{eqnarray}
\rho^{\prime} &=&  \sum_{m=0}^{N} \sum_{n=0}^{N}\int \sqrt{C_{N}^{m}C_{N}^{n}}[\sin(\theta/2)]^{m+n}e^{-i (m-n) \varphi} [\cos(\theta/2)]^{2N-m-n} \frac{ \sin(\theta)d\theta d\varphi} {4 \pi}  \vert m \rangle  \langle n \vert   \nonumber \\
&=&  \sum_{n=0}^{N}\int_{0}^{\pi} C_{N}^{n}[\sin^{2}(\theta/2)]^{n} [\cos^{2}(\theta/2)]^{N-n} \frac{ \sin(\theta)d\theta} {2}  \vert n \rangle  \langle n \vert   \nonumber \\
&=&  \sum_{n=0}^{N}2 C_{N}^{n} \int_{0}^{\pi/2} [\sin^{2}(x)]^{n+1/2} [\cos^{2}(x)]^{N-n+1/2}dx  \vert n \rangle  \langle n \vert   \nonumber \\
&=&  \sum_{n=0}^{N} C_{N}^{n} B(n+1,N-n+1)\vert n \rangle  \langle n \vert
\end{eqnarray} 
where $ B(x,y)$ is the Beta function. The ensemble average of the density matrix of the spin coherent state is not equal to the Boltzman distribution $e^{-\beta H}$ at infinite temperatures. This is mainly due to the fact that the  averaged density matrix of spin coherent state is block diagonal, whereas  the Boltzman distribution is diagonal.

We further discuss  the case where  the bath is a mixed state of the form:  the initial state $ \rho _{0}=\vert \uparrow  \rangle \otimes \rho_{b}$with a maximum  mixed state 
\begin{eqnarray}
\rho_{b}=\otimes_{j=1}^{N} [(1+p)/2\vert +  \rangle \langle +  \vert_{j}+(1-p)/2 \vert  - \rangle \langle - \vert_{j}]. 
\end{eqnarray}
where the parameter $p \in [-1,1]$, so that the polarization along $x$-axis  $P=Np$ for bath spins. The state of system evolves in  time $\rho(t)=\exp(-iHt) \rho _{0}\exp(iHt)$ and thus the spin polarization of central spin is given by $S_{0}^{z}(t)=\Tr[s_{0}^{z}\rho(t)]$. From this result, we observe that there is no quantum collapse and revival for a fully mixed initial state  $ \rho _{0}$ of the bath( Boltzman distribution with $\beta=0$) . As shown in  Fig.~\ref{QCR-random}, the larger  is polarization along of bath $x$ direction, the better collapse and revival of the central spin polarization becomes.

 \begin{figure}
\begin{center}
\includegraphics[scale=0.45]{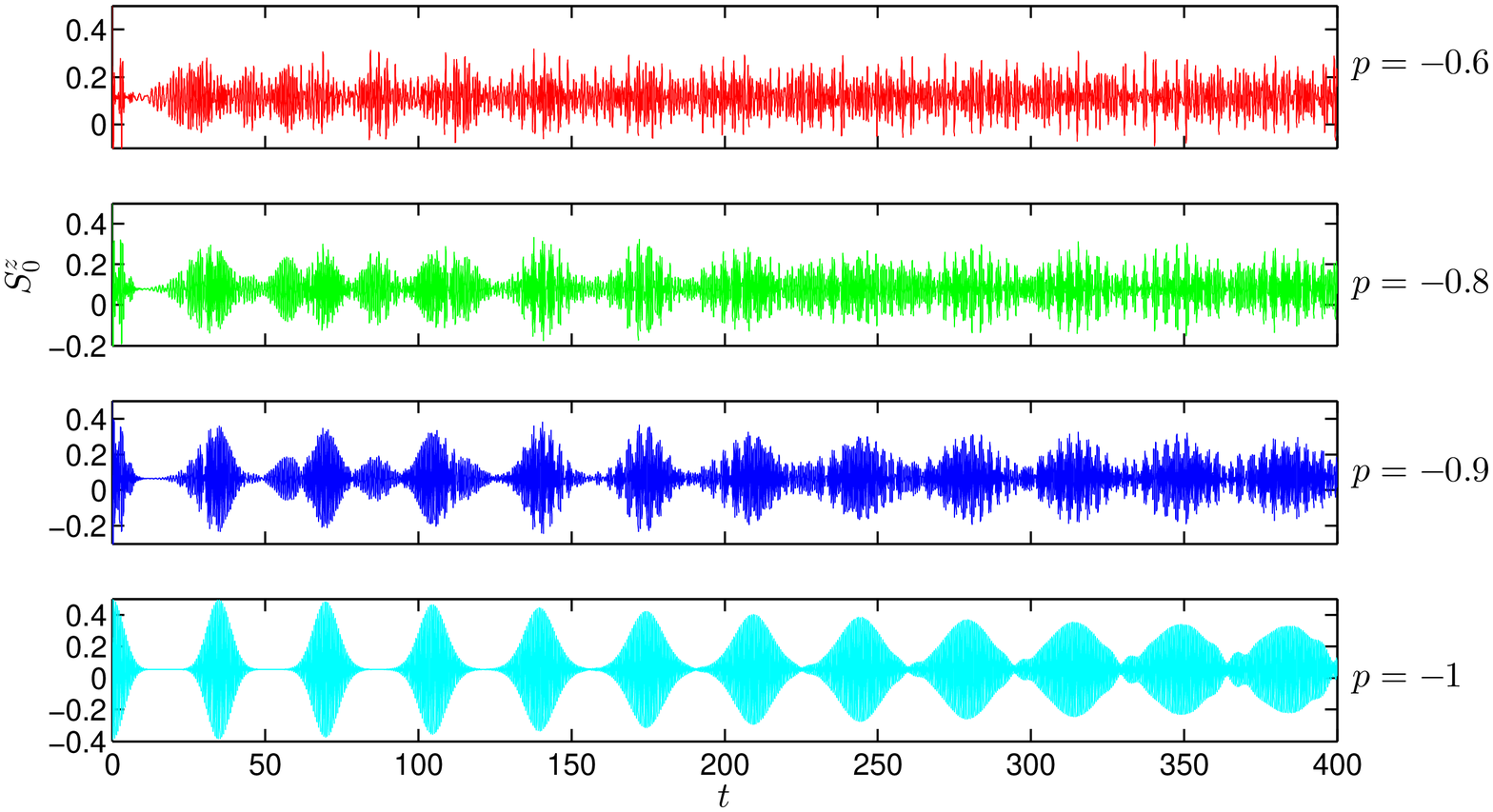}
\end{center} 
\caption{  The central spin polarization evolves in  time for the initial state $ \rho _{0}=\vert \uparrow  \rangle \otimes \rho_{b}$, where the bath spins are in different mixed state $\rho_{b}=\otimes_{j=1}^{N} [(1+p)/2\vert +  \rangle \langle +  \vert_{j}+(1-p)/2 \vert  - \rangle \langle - \vert_{j}] $ with parameter $p \in [-1,1]$.  The revival and collapse does not exist for a fully mixed state of the bath. Large polarization along the $x$-axis endow system with collapse and revival. In this figure we set $N=10, A=\Delta=B=1$.}
\label{QCR-random}
\end{figure}

\bigskip
 \noindent \emph{Reduced density matrix}. 

According to  Eq.~(\ref{Qcrsmrho}),  the reduced density matrix of  the central spin  is obtained by tracing out the degrees of freedom of bath spins. It reads 
\begin{equation}
 \rho_{cs}=\left(
\begin{array}{cc}
A(t) & B(t) \\
C(t) & D(t) \\
\end{array}
\right),
\end{equation}
 where the four matrix elements are given by 
\begin{align*}
&A(t)=\sum_{n=0}^{N}{ C_{N}^{n}}[\sin^{2}(\theta/2)]^{n}[\cos^{2}(\theta/2)]^{N-n} \vert  P_{\uparrow}^{n} \vert^2, \\
&D(t)=\sum_{n=0}^{N}{ C_{N}^{n}} [\sin^{2}(\theta/2)]^{n}[\cos^{2}(\theta/2)]^{N-n}\vert  P_{\downarrow}^{n} \vert^2, \\
&B(t)=\sum_{n=0}^{N}\sqrt{ C_{N}^{n+1}C_{N}^{n}} [\sin^{2}(\theta/2)]^{n+\frac{1}{2}}[\cos^{2}(\theta/2)]^{N-n-\frac{1}{2}} P_{\uparrow}^{n+1} (P_{\downarrow}^{n})^{*}, \\
&C(t)=B(t) ^{*}. 
\end{align*}
By diagonalizing the reduced density matrix, we obtain the eigenvalues of  $ \rho_{cs}$ 
\begin{equation}
\Lambda_{1,2}=\frac{1}{2} \pm \frac{1}{2}\sqrt{(A-D)^2+4BC}.
\end{equation}
The quantum purity and Von Neumann entropy are immediately found from $\rho_{cs}$ 
\begin{align*}
& \gamma\equiv \Tr[\rho_{cs}^2]=\frac{1}{2} + \frac{1}{2}[(A-D)^2+4BC],\\
& S(\rho_{cs})\equiv-\Tr[\rho_{cs} \ln\rho_{cs}]=-\Lambda_{1}\ln \Lambda_{1}-\Lambda_{2}\ln \Lambda_{2},
\end{align*} 
\color{black}
and a detailed discussion of their time dependence is given in the main text.

\begin{figure}
  \begin{center}
%\begin{flushleft}
\includegraphics[scale=0.4]{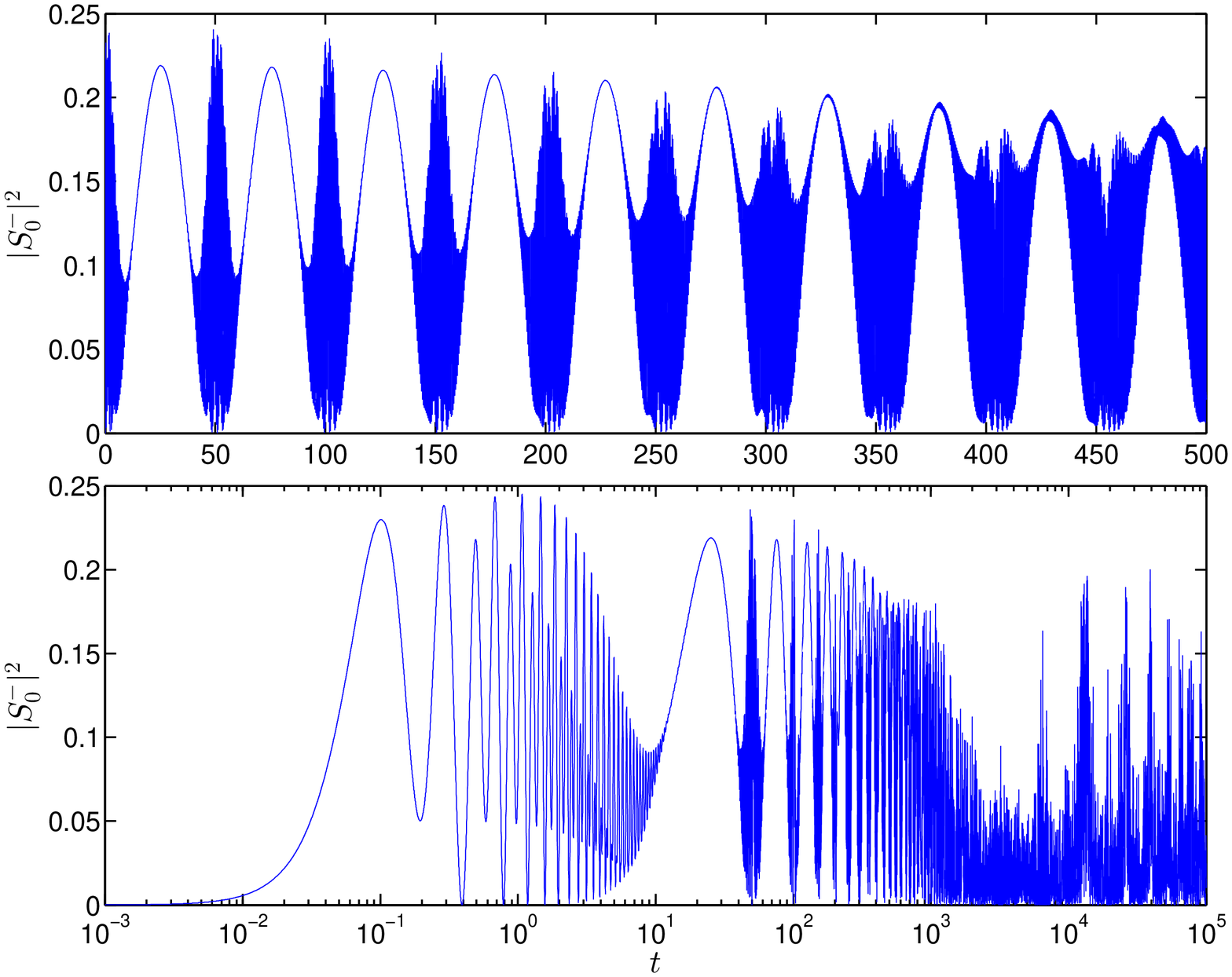}
%\end{flushleft}
\end{center} 
\caption{The evolution of coherence factor of central spin for a short time(upper panel) and a  long time (lower panel) with logarithmical scale in horizontal axis. The parameters: bath size $N=15$, the magnetic field and coupling $B=A=\Delta=1$.}
\label{Qcrsmfig5}
\end{figure}

\bigskip
\noindent\emph{Coherence factor}.

The coherence factor is defined as $ S^{-}_{0}(t) =\langle \psi(t) \vert  \mathbf{s}^{-}_{0}  \vert \psi(t) \rangle$.  It  can be also written as $S^{-}_{0}(t) =\Tr[\rho_{cs} {\bf s}^{-}_{0}]$,  i.e.,  the square norm of  the coherence factor is simply given by the  off-diagonal element of the reduced density matrix $\rho_{cs}$
\begin{equation}
\vert S^{-}_{0}(t) \vert^{2}=\vert B(t)\vert^{2}.
\end{equation}
 A plot of the time dependence of $\vert S^{-}_{0}(t)\vert^{2}$ is shown in  Fig.~\ref{Qcrsmfig5}. 
We observe that the evolution of the coherence factor  shows  large regular revivals at  short time.  For the  given initial  state, the coherence factor nearly reaches  a maximum value of  $0.25$, which reflects the large similarity of the central spin state with $\vert\phi \rangle=\frac{1}{\sqrt{2}}(\vert \uparrow \rangle_{0}+e^{-i\phi}\vert \downarrow \rangle_{0})$ (see  the discussion of fidelity below) near the middle point of collapse region. After a  long time evolution, the  coherence factor randomly oscillates,  due to  dephasing  induced by  the Rabi oscillation terms $\cos(\Omega_{n+1}t) $ .

\bigskip
 \noindent \emph{Fidelity}. 

We study the fidelity of the reduced density matrix of the 
central spin with the state $\vert\phi \rangle=\frac{1}{\sqrt{2}}(\vert \uparrow \rangle_{0}+e^{-i\phi}\vert \downarrow \rangle_{0})$, which is defined as  follows
\begin{equation}
F(\rho_{c,s},\rho_{\phi})=\left[ \Tr\sqrt{\sqrt{\rho_{\phi}} \rho_{c,s} \sqrt{\rho_{\phi}}}   \right].
\end{equation}
 Since $\rho_{\phi}=\vert\phi \rangle \langle\phi \vert$ is a pure state,  we simplify  the  above formula  to  $F(\rho_{c,s},\rho_{\phi})=\sqrt{\langle \phi \vert \rho_{c,s} \vert\phi \rangle}$. Thus, the fidelity reads  
\begin{equation}
F=\sqrt{[1+Be^{-i\phi}+Ce^{i\phi}]/2}.
\end{equation}
The features of the fidelity were discussed in connection with  Fig.~4 of our main  paper, where we observed that  in the middle  of the collapse regime, the fidelity of the  central spin against the state $\vert \phi \rangle$ alternatively takes the maximum ($>0.9$) and minimum ($<0.2$) values for the phase $\phi=\frac{\pi}{2}$ and $\phi=\frac{3\pi}{2}$, respectively. The dependence of the maximum achievable fidelity on the initial state angle of the bath is explored in Fig.\ref{Qcrsm6-F}.

%{\color{blue} Our result presents 
%the central spin precessing around an effective field along $x$, which
%is produced by the large spin with orientation $\theta = \pi/2$.
%However, the semiclassical interpretation of this central spin precessing  does not seem to reflect accurately the quantum dynamics of this model. A first observation is that a large value of the fidelity with a specific central-spin state implies a small entanglement with the bath (since the central-spin state is almost pure). Indeed, as already reported in Fig. 3 in the main text, the value of the entanglement entropy S has a pronounced minimum between revivals, when the fidelity at $ \phi = \pi/2 $ or $\phi = 3\pi/2$  is large. However, at other times there is significant entanglement, suggesting that the picture of the central spin precessing around the large spin is somewhat problematic, see Fig.~\ref{Qcrsm6-F}.
%Interestingly, the maximum fidelity always occurs at $\phi = \pi/2 $ or $ \phi = 3\pi/2$. The effect of changing $\theta$  is to corrupt the maximum achievable fidelity. We have reported this behavior in the main text. 
%}

 \begin{figure}
\begin{center}
\includegraphics[scale=0.4]{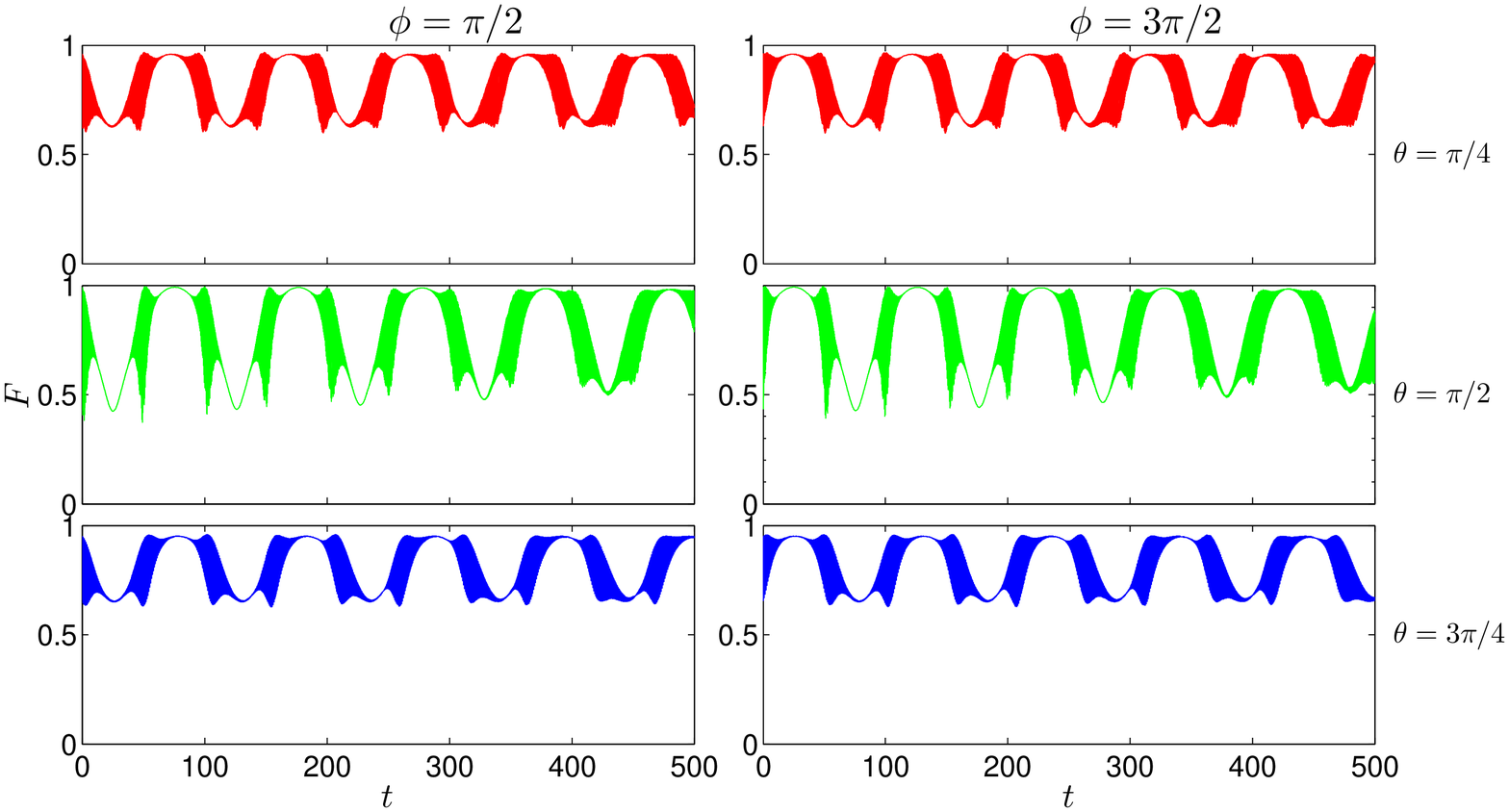}
\end{center} 
\caption{The  time evolution of the fidelity of the reduced density matrix of the central spin $\rho_{cs}$ against the
state  $\vert \phi \rangle=\frac{1}{\sqrt{2}}[\vert \uparrow \rangle_{0}+e^{-i\phi}\vert \downarrow \rangle_{0}]$ , here $\phi=\pi/2, \,3\pi/2 $,  for three different initial state angles of bath $\theta$, whose maximum value of fidelity are respectively $0.937, \,0.987, \,0.919$. The parameters take $A=\Delta=B=1, N=12$. }
\label{Qcrsm6-F}
\end{figure}

\bigskip
 \noindent \emph{Correlation function}.

\begin{figure}
  \begin{center}
\includegraphics[scale=0.35]{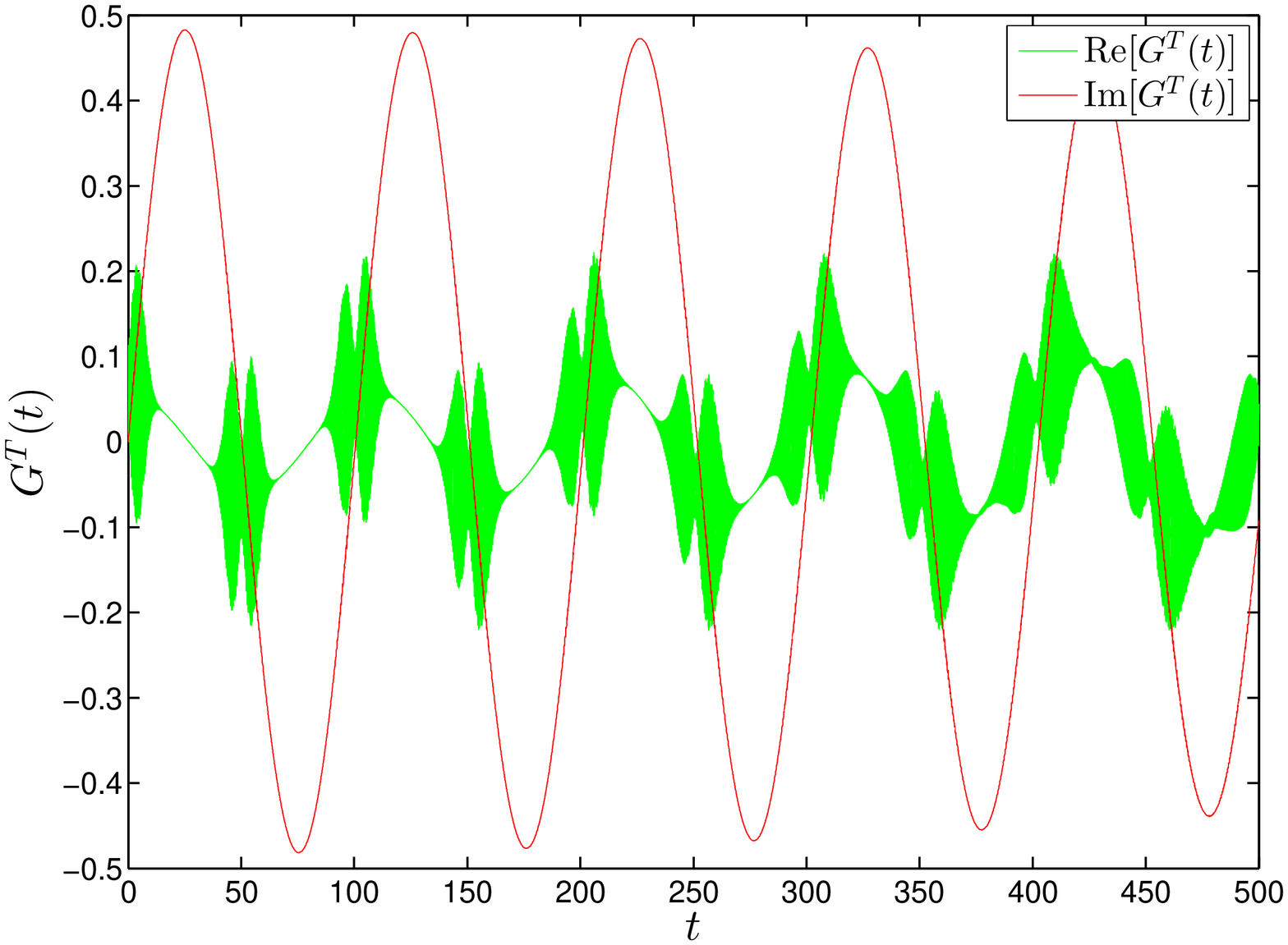}
\end{center} 
\caption{The time-dependent transverse correlation function  $G^{T}(t)$  between  the  central spin and bath spins. Green line:  the real part of correlation function; Red line:  the imaginary part of correlation fucntion. Bath size $N=15$, the magnetic field and coupling are $B=A=\Delta=1$.}
\label{ch4Qcr_correlation}
\end{figure}

 The  longitudinal correlation function is defined
\begin{equation}
G^{z}(t)=\langle \psi(t) \vert \mathbf{s}^{z}_{0} \mathbf{J}^{z} \vert \psi(t) \rangle.
\end{equation}
By using the wave function (\ref{Qcrwf}),  we derive the longitudinal correlation function as follows 
\begin{align}
&G^{z}(t)=\frac{1}{2}\sum_{n=0}^{N}C_{N}^{n}[\sin^{2}(\theta/2)]^{n}[\cos^{2}(\theta/2)]^{N-n}  \left[ (n-\frac{N}{2}) \vert P_{\uparrow}^{n}\vert^2-(n+1-\frac{N}{2}) \vert P_{\downarrow}^{n}\vert^2 \right].
\end{align}
After substituting the $P_{\uparrow}^{n}, P_{\downarrow}^{n}$ and simplifying   the  above formula, we can easily get the correlation function
\begin{align}
G^{z}(t)=&\frac{1}{2}\sum_{n=0}^{N}C_{N}^{n}[\sin^{2}(\theta/2)]^{n}[\cos^{2}(\theta/2)]^{N-n}  \left[ \frac{(n-\frac{N}{2}) \Delta^{2}_{n+1}-2b_{n+1}A^{2}}{(\Omega_{n+1})^2} \right. \nonumber \\
& \left. +  (n-\frac{N}{2}+\frac{1}{2})\frac{4b_{n+1}A^2}{(\Omega_{n+1})^2} \cos(\Omega_{n+1}t) \right].
\end{align}
The time-dependent transverse correlation function is defined by 
\begin{equation}
G^{T}(t)=\langle  \Phi_{0} \vert \mathbf{J}^{+}(t){\bf s}^{-}_{0}  \vert \Phi_{0} \rangle,
\end{equation}
where $\mathbf{J^{+}}(t)= e^{iHt} \mathbf{J^{+}} e^{-iHt}$. For the  initial state $\vert \Phi_{0} \rangle$, the transverse correlation function also equals  to
\begin{equation*}
G^{T}(t)=\langle \psi_{\uparrow}(t) \vert \mathbf{J^{+}}\vert \psi_{\downarrow}(t) \rangle.
\end{equation*}
Here the two wave  functions are  $\vert \psi_{\uparrow}(t) \rangle=e^{-iHt}\vert \uparrow \rangle_{0} \vert \theta \rangle $ and $\vert \psi_{\downarrow}(t) \rangle=e^{-iHt}\vert \downarrow \rangle_{0} \vert \theta \rangle $, which have been obtained in previous parts  of this  supplementary   material. After substituting the wave  functions into the above formula, the correlation function is obtained  as 
\begin{align}
G^{T}(t)=&\frac{1}{2}\sum_{n=0}^{N}C_{N}^{n}[\sin^{2}(\theta/2)]^{n}[\cos^{2}(\theta/2)]^{N-n} \left\lbrace  \frac{2b_{n}A\Delta_{n+1}-2b_{n+1}A \Delta_{n}}{\Omega_{n} \Omega_{n+1}} \sin(\frac{\Omega_{n}}{2}t) \sin(\frac{\Omega_{n+1}}{2}t) \right.  \nonumber \\
& \left.  + i \left[  \frac{2b_{n+1}A}{\Omega_{n+1}} \sin(\frac{\Omega_{n+1}}{2}t)\cos(\frac{\Omega_{n}}{2}t)- \frac{2b_{n}A}{\Omega_{n}} \sin(\frac{\Omega_{n}}{2}t)\cos(\frac{\Omega_{n+1}}{2}t) \right]  \right\rbrace. 
\end{align}
 The  correlation function $G^{T}(t)$ measures   the probability of the  bath spins flipping up at time $t$ when  the central spin flips down at the  initial time $t=0$. A plot of $G^{T}(t)$ is displayed in Fig.~\ref{ch4Qcr_correlation}, showing the oscillatory nature of its time evolution.

\bigskip
\noindent\emph{Loschmidt echo}. 

The Loschmidt echo is defined as $L(t)=\vert \langle \Phi_{0}\vert \psi(t) \rangle \vert^2$. After substituting the initial state $\vert \Phi_{0}  \rangle$ and  the  wave function $\vert \psi(t)\rangle=e^{-iHt} \vert \Phi_{0}  \rangle$ into  the  definition, we obtain  $L(t)$ in terms of  the overlap $g(t)=\langle \Phi_{0}\vert \psi(t) \rangle$,  given by 
\begin{equation}
g(t)=\sum_{n=0}^{N}C_{N}^{n}[\sin^{2}(\theta/2)]^{n}[\cos^{2}(\theta/2)]^{N-n} P_{\uparrow}^{n},
\end{equation}
 where $P_{\uparrow}^{n}= -i\frac{\Delta_{n+1}}{ \Omega_{n+1}}\sin( \frac{\Omega_{n+1}t}{2})+\cos( \frac{\Omega_{n+1}t}{2})$. The Loschmidt echo is obtained by taking the square norm of the overlap $L(t)=\vert g(t) \vert^2$. An example of time dependence is shown in Fig.~\ref{QcrsmLecho}.\
 
The collapse and revival is robust against the inhomogeneity with the coupling.  Seeing Fig.~\ref{inLecho}, there is still collapse and revival in rather long time ($t \sim  200$) at weak inhomogeneity  $ \alpha=1$.

\begin{figure}
  \begin{center}
\includegraphics[scale=0.4]{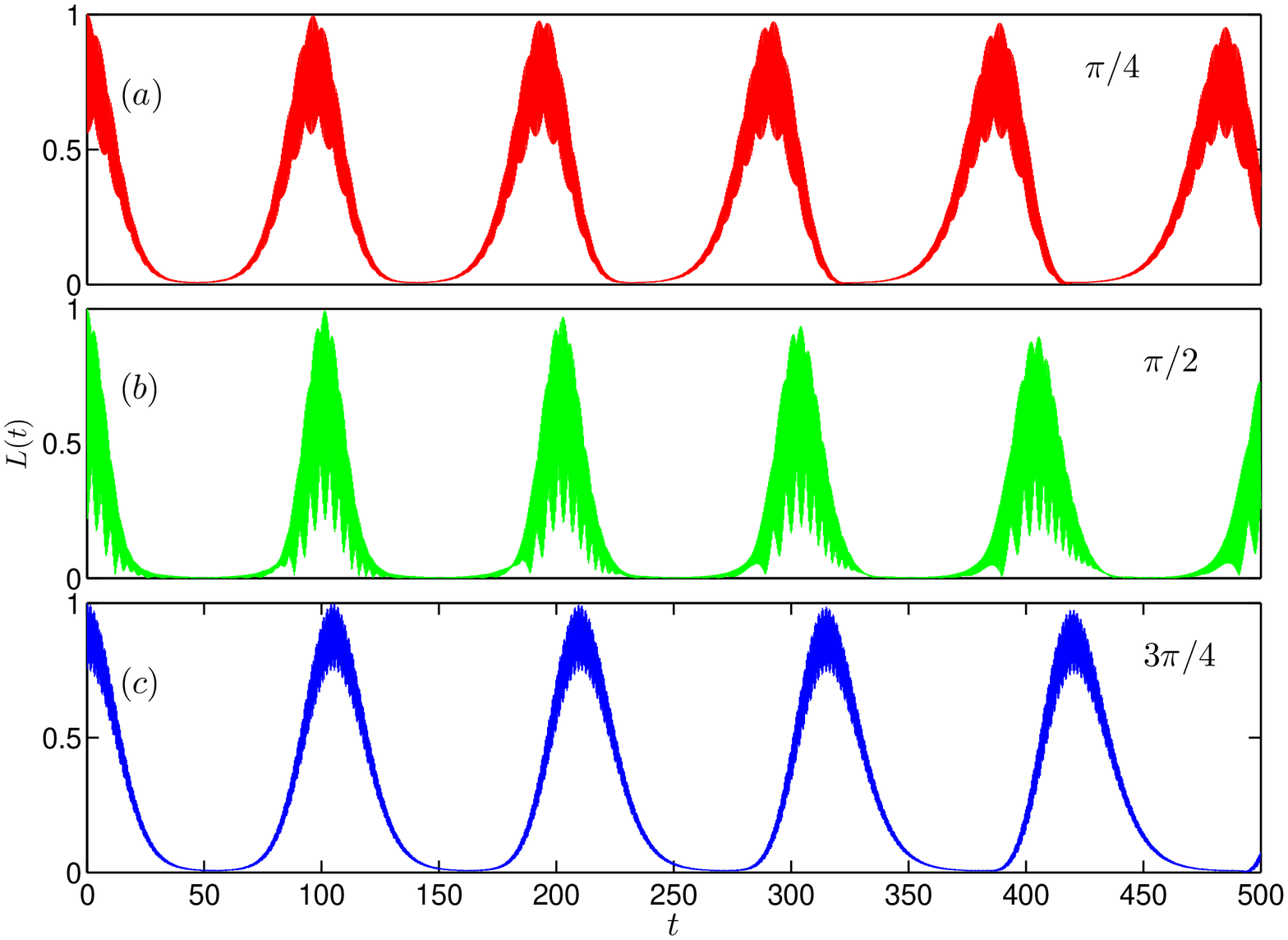}
\end{center} 
\caption{The Loschmidt echo vs time  for different initial state angles.  (a): $\theta=\pi/4$; (b): $\theta=\pi/2$; (c): $\theta=3\pi/4$.  Bath size $N=15$. The magnetic field and coupling  are  $B=A=\Delta=1$.}
\label{QcrsmLecho}
\end{figure} 
 
 \begin{figure}
  \begin{center}
\includegraphics[scale=0.4]{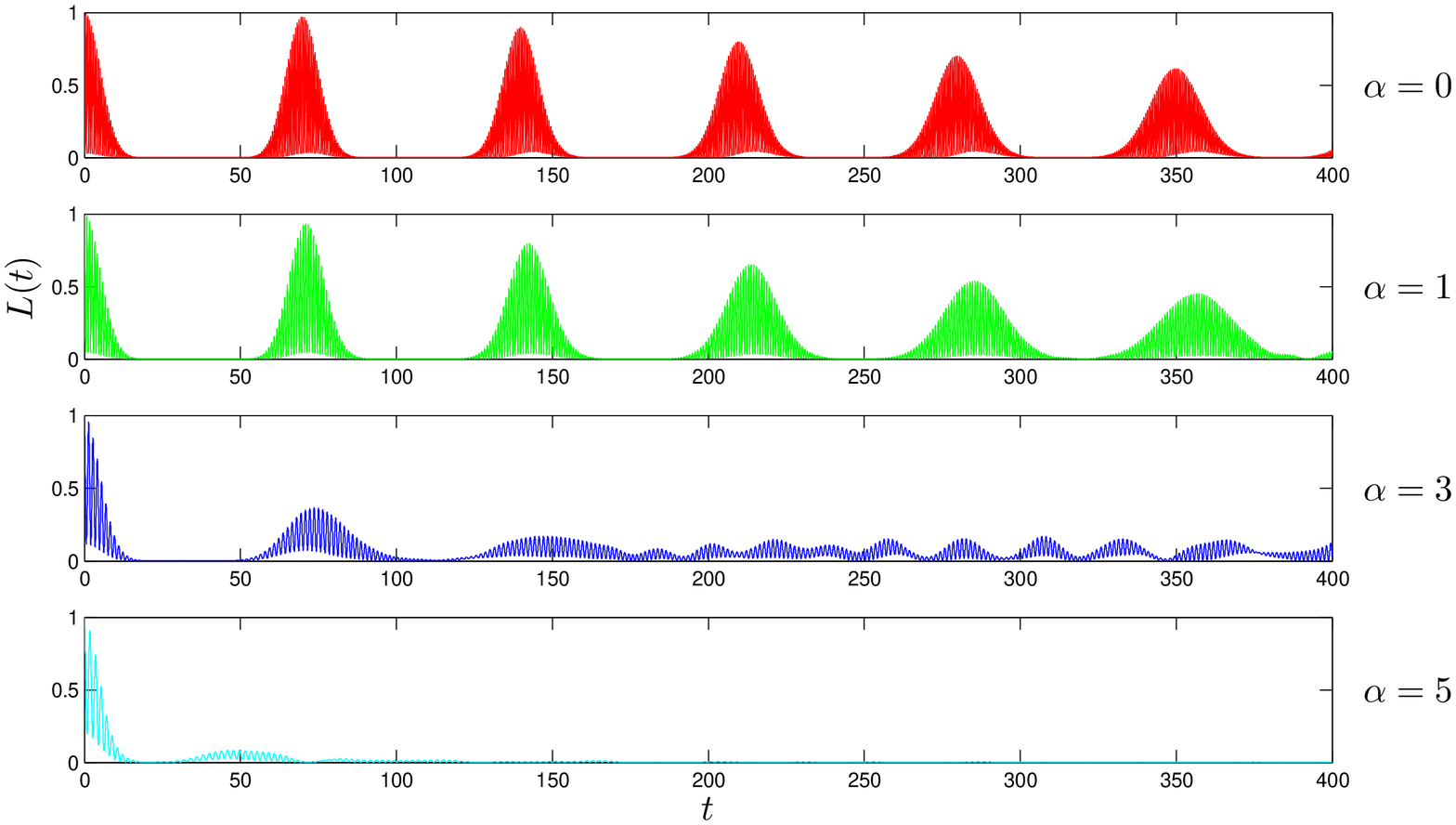}
\end{center} 
\caption{The Loschmidt echo vs time  for different inhomogeneity factor $\alpha$.  Bath size $N=10$. The magnetic field and coupling  are  $B=A=1$.}
\label{inLecho}
\end{figure}

{
\bigskip
 \noindent \emph{Connection to the Jaynes-Cummings model. }
 
 Through the  Holstein-Primakoff transformation, we built up in the main text  the connection between the Hamiltonian of central spin model Eq. (3) to the  Jaynes-Cummings model at a large $N$ limit. This  connection reveals a statistical nature of Holstein-Primakoff transformation. 
 In fact, such a statistical connection  demands a special choice of initial state in  the central spin model.  Here the Figure~\ref{Qcrsm8} quantitatively shows  how well the quantum collapse and revival dynamics  of central spin model with a large number of bath spins  simulates the one of   the Jaynes-Cummings model with a  certain average photon  number.

\begin{figure}
  \begin{center}
%\begin{flushleft}
\includegraphics[scale=0.35]{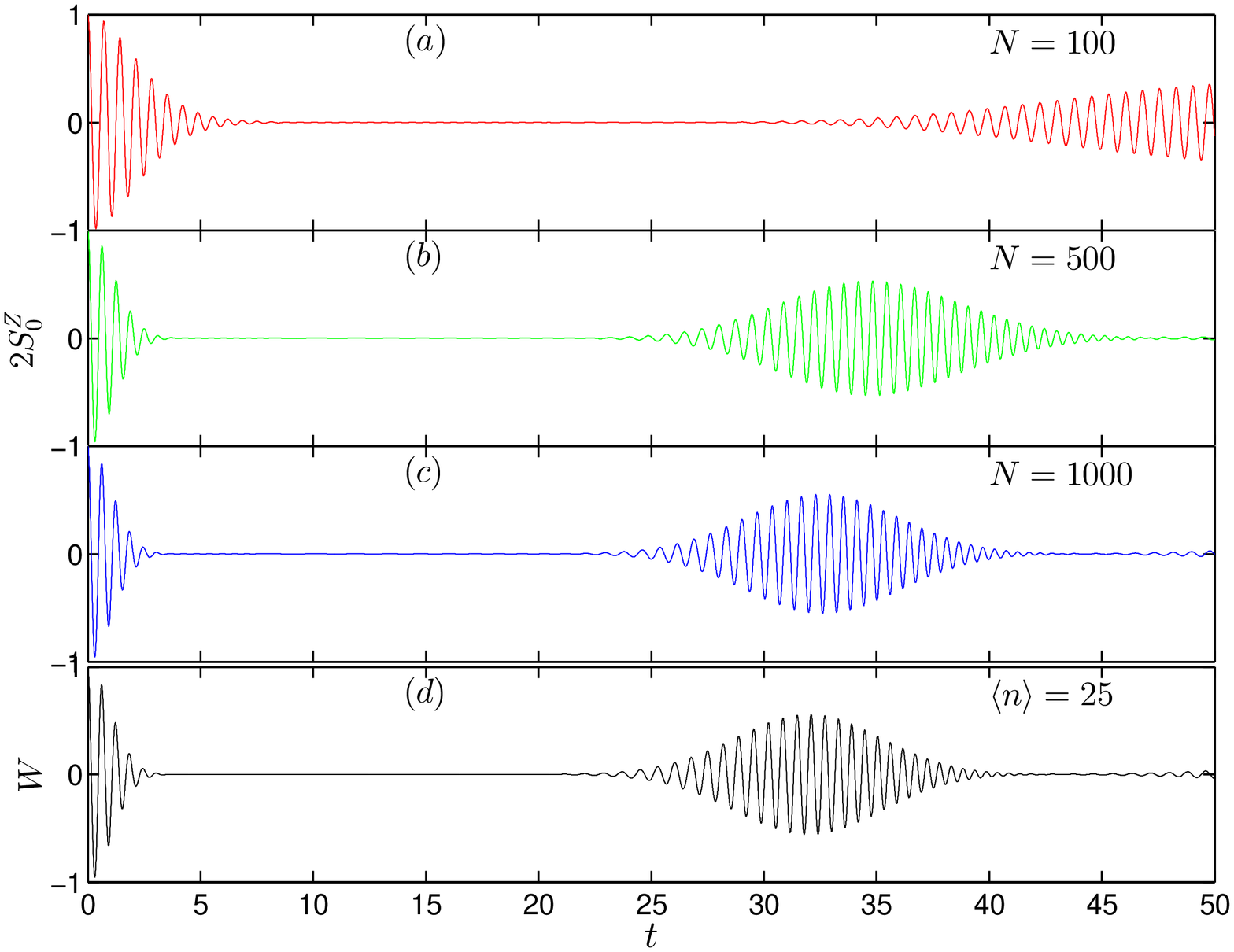}
%\includegraphics[scale=0.35]{Qcrfig4.eps}
%\end{flushleft}
\end{center} 
%\begin{widetext}
\caption{Statistical mapping in Holstein-Primakoff transformation. (a),  (b),   and (c) show the evolutions of  $2S_0^z$  at different bath sizes $N=100,500,1000$. Here  we take  detuning $\Delta_{n}=0$ (namely,  $B=\Delta=0$), transverse coupling $A=1/\sqrt{N}$, initial state angle satisfying $N\sin^{2}(\theta/2)=25$. (d) shows the inversion of the Jaynes-Cummings model with  detuning $\Delta_{JC}=0$  and average photon  number $ \langle n \rangle=25$, see  Ref.~\cite{Scully:1997}. We observe  that  the two figures (c) and (d)  are almost  identical, reflecting  a statistical mapping: from a fermionic  field  to a bosonic field.  }
\label{Qcrsm8}
%\end{widetext}
\end{figure}

}

\bigskip
\noindent\emph{Eigenstates}.

Firstly, we assume  that  the eigenstate is  a general  superposition of all  basis states 
\begin{equation}
\vert u \rangle=\sum_{n=0}^{N} \alpha_{n} \vert \uparrow \rangle_{0} \vert n \rangle+\sum_{n=0}^{N} \beta_{n} \vert \downarrow \rangle_{0} \vert n \rangle.
\end{equation}
The eigen-equation reads
\begin{equation}
H \vert u \rangle=E \vert u \rangle.
\label{eigE}
\end{equation}
 where the action of $H$ is given by 
\begin{align*}
&H \vert \uparrow \rangle_{0} \vert n \rangle= w_{n} \vert \uparrow \rangle_{0} \vert n \rangle +\sqrt{b_{n+1}}A  \vert \downarrow \rangle_{0}\vert n+1 \rangle, \\
&H \vert \downarrow \rangle_{0}\vert n \rangle=  -w_{n} \vert \downarrow \rangle_{0}\vert n \rangle +\sqrt{b_{n}}A\vert \uparrow \rangle_{0} \vert n-1 \rangle.
\end{align*}
Comparing the amplitudes on both sides of Eq.~(\ref{eigE}) immediately gives: 
\begin{align}
& (E-w_{n}) \alpha_{n} =\sqrt{b_{n+1}}A   \beta_{n+1}, \\
& (E+w_{n}) \beta_{n} =\sqrt{b_{n}}A   \alpha_{n-1}.
\end{align}
We can obtain the eigenergy from the above two equations
\begin{equation}
E^{2}+(w_{n+1}-w_{n})E-[ w_{n}w_{n+1}+ b_{n+1}A^2] =0,
\end{equation}
namely the eigenergy reads
\begin{align*}
E_{n}^{1,2}=\frac{(w_{n}-w_{n+1}) \pm \sqrt{(w_{n}+w_{n+1})^2+4b_{n+1}A^2}}{2}\color{red} , \color{black}
\end{align*}
which actually are same  as  $\lambda_{1,2}$ in Eq.(7).
In fact,   since the magnization is a conserved quantity ($[H,\mathbf{s}_{0}^{z}+\mathbf{J}^{z}]=0$),  we can decompose the Hilbert space according to  its eigenvalue. For  the  subspace with $n$ flipped bath spins, the relevant subspaces are given by 
\begin{align}
& \vert \uparrow \rangle_{0} \vert n \rangle \rightarrow \vert \downarrow \rangle_{0}\vert n+1 \rangle, \\
& \vert \downarrow \rangle_{0}\vert n \rangle   \rightarrow   \vert \uparrow \rangle_{0} \vert n-1 \rangle,
\end{align}
which  are in strict analogy to the Jaynes-Cummings model, where the excitation number is conserved. Therefore, 
 we obtain the  following eigenfunctions
\begin{align}
&\vert u^{+}_{n}  \rangle=  \vert \uparrow \rangle_{0} \vert n \rangle+ \frac{(w_{n}-w_{n+1}) + \Omega_{n+1}}{2\sqrt{b_{n+1}}A} \vert \downarrow \rangle_{0} \vert n+1 \rangle, \\
&\vert u^{-}_{n} \rangle=  \vert \uparrow \rangle_{0} \vert n \rangle+ \frac{(w_{n}-w_{n+1}) - \Omega_{n+1}}{2\sqrt{b_{n+1}}A}  \vert \downarrow \rangle_{0} \vert n+1 \rangle.
\end{align}
After normalization, the eigenstates read \cite{chesi}
\begin{align}
& \vert u^{+}_{n}  \rangle=  \cos(\theta_{n}) \vert \uparrow \rangle_{0} \vert n \rangle+ \sin(\theta_{n}) \vert \downarrow \rangle_{0} \vert n+1 \rangle, \\
& \vert u^{-}_{n} \rangle=  \sin(\theta_{n}) \vert \uparrow \rangle_{0} \vert n \rangle- \cos(\theta_{n}) \vert \downarrow \rangle_{0} \vert n+1 \rangle,
\end{align}
here the angle  satisfies  $\tan(\theta_{n})=\sqrt{\frac{ \Omega_{n+1}+\Delta_{n+1}}{\Omega_{n+1}-\Delta_{n+1}}}$.

 \section{ Inhomogeneous case}

%For inhomogeneous case, the state basises $ \vert \uparrow \rangle  \vert n \rangle $ and $\vert \downarrow \rangle  \vert n \rangle$ is not suitable basis for the eigenfunction of inhomogeneous central spin. In fact the eigenstate basises of the inhomogeneous central spin model, is given as following by Bethe ansatz
%\begin{equation}
%\vert \nu_{1},\cdots,\nu_{M} \rangle=\prod_{\alpha=1}^{M}B_{\nu_{\alpha}} \vert \Uparrow \rangle=\prod_{\alpha=1}^{M}\sum_{j} \frac{s_{j}^{-}}{\nu_{\alpha}-\epsilon_{j}}  \vert \Uparrow \rangle.
%\end{equation}

 \subsection{Bethe ansatz solution}
 The central spin model describes  a  spin coupled to bath spins via  a long-range interaction, whose  Hamiltonian is written as 
\begin{equation}
H=B\mathbf{s} ^{z}_{0}+2 \sum_{j=1}^{N}A_{j}\textbf{s}_{0} \cdot\textbf{s}_{j},
\label{Hami}
\end{equation}
here  the subindex ``0" labels the site of central spin,  whereas the subindices $1 \rightarrow N$  label the sites of the bath spins. For our convenience in the following discussion, we introduce the anisotropic coupling  parameters $A_{j}=1/(\epsilon_{0}-\epsilon_{j})$ with $j=1, \ldots, M$,  the magnetic field $B=-{2}/{g}$ and central spin energy level $ \epsilon_{0}=0$. 
Using   the  algebraic Bethe ansatz,  we can obtain the the eigenfunction of the Hamiltonian (\ref{Hami}) with   $M$ down spins  \cite{J.Dukelsky2004,H.-Q. Zhou}
\begin{equation}
\vert \nu_{1},\cdots,\nu_{M} \rangle=\prod_{\alpha=1}^{M}B_{\nu_{\alpha}} \vert \Uparrow \rangle=\prod_{\alpha=1}^{M}\sum_{j} \frac{s_{j}^{-}}{\nu_{\alpha}-\epsilon_{j}}  \vert \Uparrow \rangle.
\label{EigF}
\end{equation}
Here we chose a fully polarized state   $\vert \Uparrow \rangle$ as the  reference state. 
There are $M$ unknown variables  $\lbrace \nu_\alpha \rbrace$ with $\alpha =1,\ldots, M$  satisfy  the  Bethe ansatz equations
\begin{equation}
 \sum \limits_{j} \frac{1}{\nu_{\alpha}-\epsilon_{j}}=\frac{2}{g}+\sum_{\beta \neq \alpha , \beta=1}^{M} \frac{2}{\nu_{\alpha}-\nu_{\beta}},\,\,\,\,\alpha =1,\ldots, M, 
 \label{BAE}
\end{equation}
which are  also called Richardson-Gaudin equations. 
 There are $C_{N+1}^{M}$ sets of solutions to Eq(\ref{BAE}) and  $C_{N+1}^{M}$ sets of the eigenfunctions  $ \vert \nu_{1},\cdots,\nu_{M} \rangle$, forming the subspaces for $M$  down spins. Moreover, eigenenergy is given by 
\begin{equation}
  E=\frac{B}{2}+\frac{1}{2} \sum_{j=1}^{N} \frac{1}{\epsilon_{0}-\epsilon_{j}}-\sum_{\alpha=1}^{M}\frac{1}{\epsilon_{0}-\nu_{\alpha}}.
  \label{Energy}
 \end{equation}

We first consider the initial state  $\vert \Phi_{0} \rangle=s_{a_{1}}^{-} \cdots s_{a_{M}}^{-} \vert \Uparrow \rangle$, where $s_{a _j}^-$ is the lowering operator acting on the reference state $\vert \Uparrow \rangle$. Here the index  $a_j$ denotes the spin flipping site,  ranging  from  $``0"$ to $``N"$. The wave function evolves in time $\vert \psi(t) \rangle= e^{-iHt}\vert \Phi_{0} \rangle$. Using the Bethe ansatz wave function, we obtain  exact evolution of  spin polarization at an  arbitrary site $j$
% \begin{widetext}
 \begin{eqnarray}
& s_{j}^{z}(t)&=\frac{1}{2}-\sum_{k}\sum_{k^{\prime}}\vert N_{\nu_{k}}\vert^{2}\vert N_{\nu_{k^{\prime}}}\vert^{2} \left[ \sum_{\mathcal{P}\in \lbrace a_{1},\cdots,a_{M} \rbrace}\frac{1}{\prod \limits_{\alpha=1}^{M}(\nu_{\alpha,k}-\epsilon_{{\mathcal{P}_{\alpha}})}} \right] \left[ \sum_{\mathcal{P}}\frac{1}{\prod \limits_{\alpha=1}^{M}(\nu_{\alpha,k^{\prime}}-\epsilon_{{\mathcal{P}_{\alpha}})}} \right] \nonumber \\
 &\sum\limits_{ j_{1}< \cdots <j_{M}} &(\sum_{\alpha}^{M}\delta_{jj_{\alpha}}) \left[ \sum_{\mathcal{Q}\in \lbrace j_{1},\cdots,j_{M} \rbrace}\frac{1}{\prod \limits_{\alpha=1}^{M}(\nu_{\alpha,k}-\epsilon_{{\mathcal{Q}_{\alpha}})}} \right] \left[ \sum_{\mathcal{Q}}\frac{1}{\prod \limits_{\alpha=1}^{M}(\nu_{\alpha,k^{\prime}}-\epsilon_{{\mathcal{Q}_{\alpha}})}} \right] \cos(w_{kk^{\prime}}t),
 \label{sjz}
\end{eqnarray}
%  \end{widetext}
where $``\mathcal{P}"$ and  $``\mathcal{Q}"$  mean summing over all permutations of indies  $\lbrace a_{1},\cdots,a_{M} \rbrace$ and $\lbrace j_{1},\cdots,j_{M} \rbrace$, respectively.  
Here the parameters $\left\{ \epsilon_j\right\}$ are introduced via the  inhomogeneous couplings $A_{j}=1/(\epsilon_{0}-\epsilon_{j})$ with a constant $\epsilon_0=0$, providing a realistic randomness of bath spins. 
Whereas $\left\{ \nu_{k}\right\}$ denote the roots of the  Bethe ansatz equations, $N_{\nu_{k}}$ is the normalization factor of the Bethe ansatz wave function.
The Rabi frequencies between different energy levels $w_{kk^{\prime}}=E_{k^{\prime}}-E_{k}$ are determined by the Bethe ansatz solution.
Coherent nature of these Rabi oscillations leads to a rich quantum dynamics \cite{He-Guan:preprint}.
Fig~ \ref{Qcrfig0} (a) and (b) show time evolution of central spin polarization, where we considered an   exponential decay of coupling amplitudes $A_{j}=A/N \exp(-j/N)$ with $j=1,\ldots, N$. 
 For different intinial states, the  central spin polarization  displays  oscillation  structure in time, revealing the  propagation of the  local information  into  bath spins. We observe that  the system does not get thermalized even   in an infinitely   long time.

  \begin{figure}
\begin{center}
\includegraphics[scale=0.35]{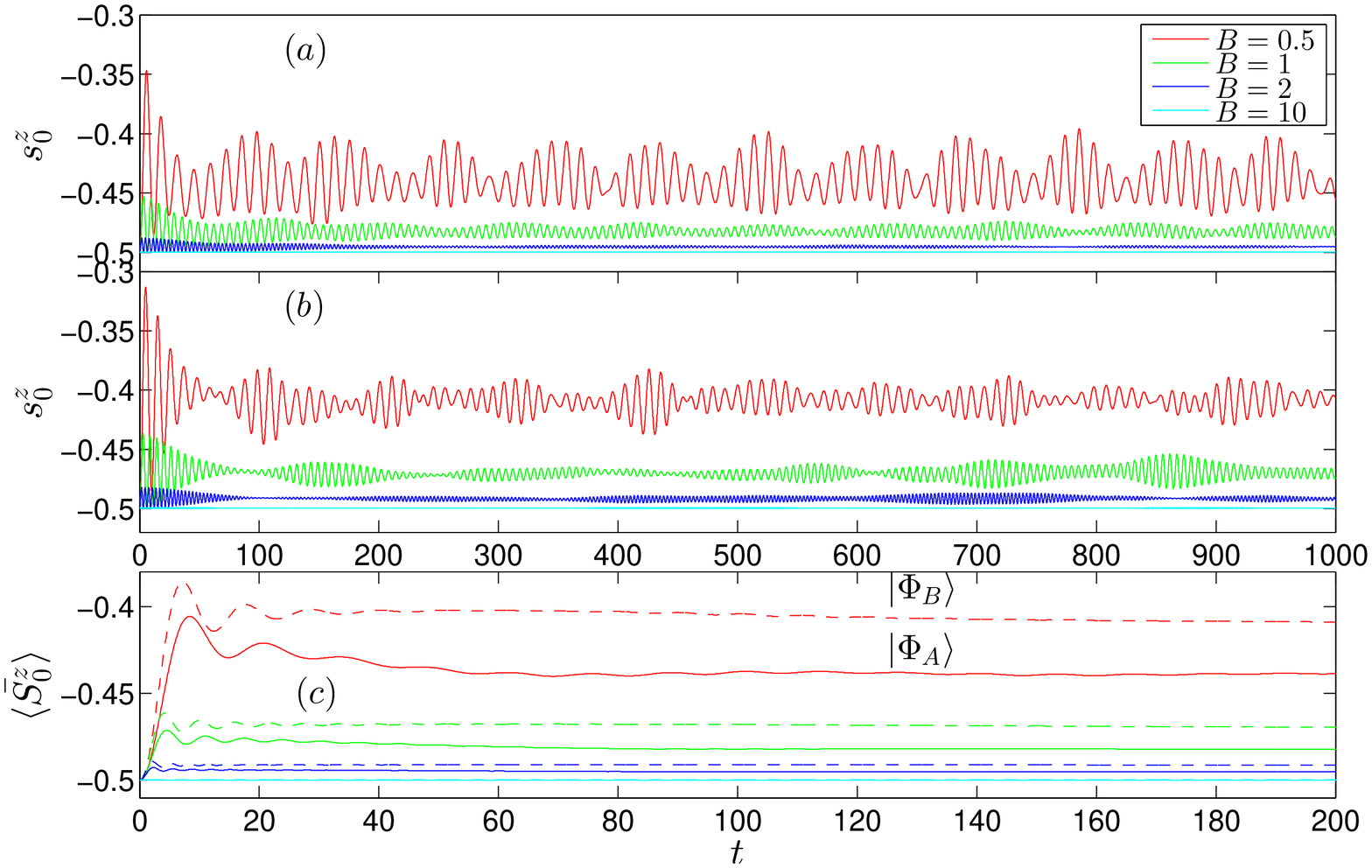}
\end{center} 
\caption{ Time  evolution of central  spin polarization for two different initial states: (a) Initial state with the locations of the  down spins at $\vert \Phi_{A} \rangle=\vert 0,1,2,3,4\rangle$. (b) Initial state with  the locations of the  down spins at $\vert \Phi_{B} \rangle=\vert 0,2,4,6,8\rangle$. Numerical calculation was carried out from the result (\ref{sjz}) with the  bath spins  $N=10$ and the number of down spins  $M=5$.  (c) Time average of central spin polarization, solid line denotes the result for the  Initial state $\vert \Phi_{A} \rangle$, dashed line denotes the result for the initial state  $\vert \Phi_{B} \rangle$. This shows no thermaliztion even in long time limit. }
\label{Qcrfig0}
\end{figure}   

 \begin{figure}
\begin{center}
\includegraphics[scale=0.45]{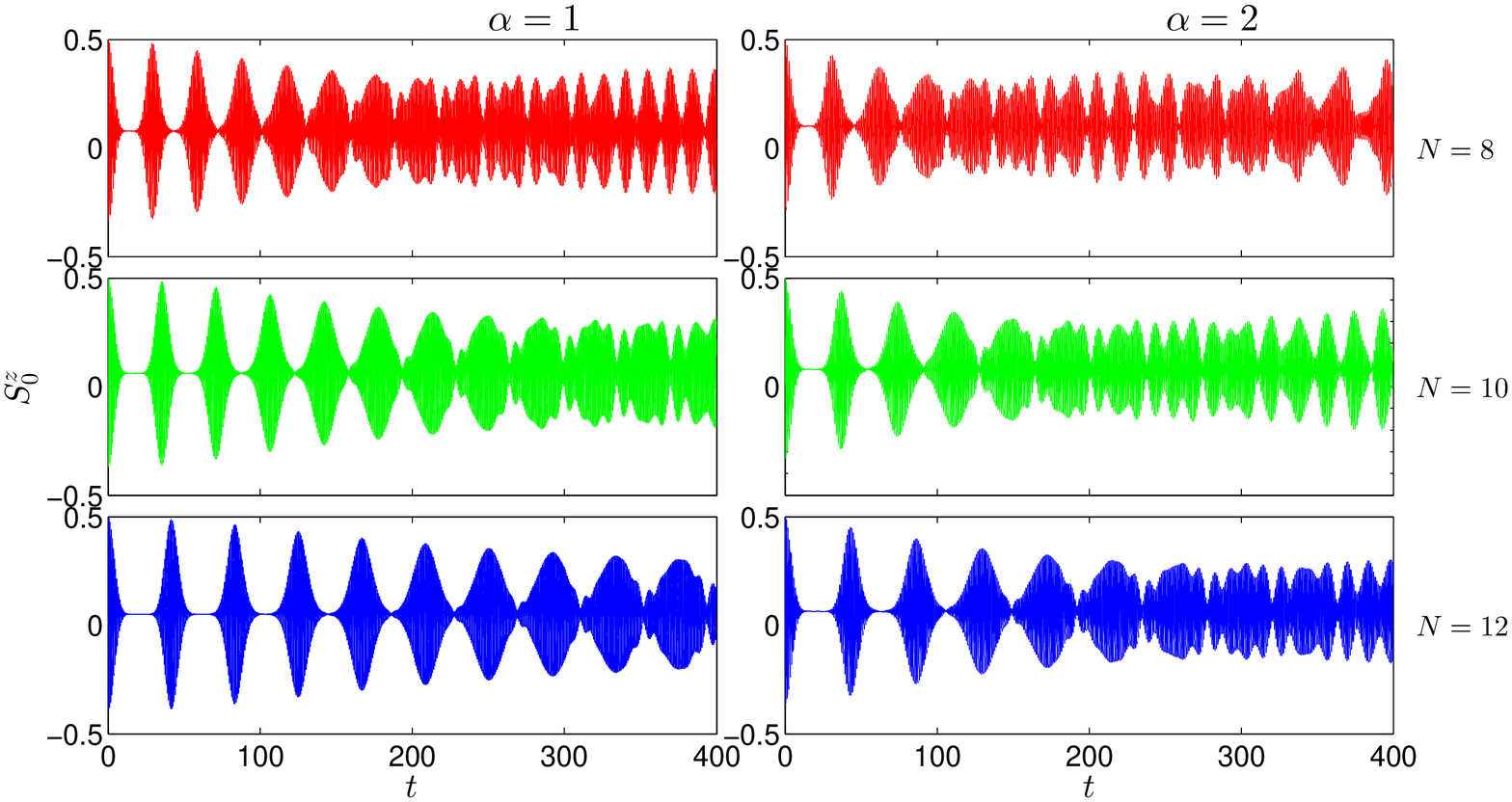}
\end{center} 
\caption{The central spin polarization evolve with time by exact diagonalization under different uniform factor $\alpha$ and bath number $N$, where the isotropic coupling take $A_{j}=A \exp(-\alpha(j-1)/N)$, $A=1,B=1,$. }
\label{inNalpha}
\end{figure}  

{
 \subsection{ Inhomogeneity effect}
In order to  characterize the effect of inhomogeneous  couplings, we first consider a  quantity, which is  called reduced bath angular momentum , 
\begin{equation}
\tilde{R}=\frac{\langle \hat{J}^{2} \rangle_{in}}{\langle \hat{J}^{2} \rangle_{ho}}=\langle \hat{J}^{2} \rangle/ \frac{N}{2}(\frac{N}{2}+1)
\end{equation}
where the bath spin operator $\hat{J}=\sum_{j}  \hat{s}_{j}$, the subindexies  "in, ho " refer to the inhomogeneous case or homogeneous case. 

We now  show the condition of existence of conserved quantity $\hat{J}^{2}$. 
The Hamiltonian
\begin{equation}
H=B\mathbf{s} ^{z}_{0}+2 \sum_{j=1}^{N}A_{j}\mathbf{s}_{0} \cdot \mathbf{s}_{j},
\end{equation}
at the same time bath spin operator $\hat{J}^{2}=\sum_{m}\sum_{n} \mathbf{s}_{m} \cdot \mathbf{s}_{n}$. We calculate the commutator $ [H,\hat{J}^{2} ]$
\begin{align*}
& [H,\hat{J}^{2} ]=2 \sum_{j}A_{j} \sum_{m}\sum_{n} [\mathbf{s}_{0} \cdot \mathbf{s}_{j},   \mathbf{s}_{m} \cdot \mathbf{s}_{n}] \\
&= \sum_{j}A_{j} \sum_{m}\sum_{n}\{  [\mathbf{s}_{0} \cdot \mathbf{s}_{j},   \mathbf{s}_{j} \cdot \mathbf{s}_{n}] \delta_{jm}+[\mathbf{s}_{0} \cdot \mathbf{s}_{j},   \mathbf{s}_{m} \cdot \mathbf{s}_{j}] \delta_{jn} \} \\
&=2 \sum_{j}A_{j} \sum_{m} \{(\mathbf{s}_{0} \cdot \mathbf{s}_{j})(\mathbf{s}_{j} \cdot \mathbf{s}_{m})- (\mathbf{s}_{j} \cdot \mathbf{s}_{m}) (\mathbf{s}_{0} \cdot \mathbf{s}_{j})\} \\
&=2 \sum_{j}A_{j} \sum_{m} \{(\mathbf{s}_{0} \cdot \mathbf{s}_{m})I+i (\mathbf{s}_{0} \times \mathbf{s}_{m})\cdot  \mathbf{s}_{j}-(\mathbf{s}_{m} \cdot \mathbf{s}_{0})I-i (\mathbf{s}_{m} \times \mathbf{s}_{0})\cdot  \mathbf{s}_{j}\} \\
& =2[  i \sum_{j}\sum_{m} A_{j}   (\mathbf{s}_{0} \times \mathbf{s}_{m})\cdot  \mathbf{s}_{j}+i \sum_{j}\sum_{m}  A_{m}  (\mathbf{s}_{0} \times \mathbf{s}_{j})\cdot  \mathbf{s}_{m}] \\
& =2[  i \sum_{j}\sum_{m} A_{j}   (\mathbf{s}_{0} \times \mathbf{s}_{m})\cdot  \mathbf{s}_{j}+i \sum_{j}\sum_{m}  A_{m}  (\mathbf{s}_{m} \times \mathbf{s}_{0})\cdot  \mathbf{s}_{j}]  \\
& =2[  i \sum_{j}\sum_{m} (A_{j}-A_{m})   (\mathbf{s}_{0} \times \mathbf{s}_{m})\cdot  \mathbf{s}_{j} ] .
\end{align*} 
Thus we see clearly that only if  $A_{j}=A_{m}$ for arbitrary $j,m$ ( homogeneous  coupling ),  then $[H,\hat{J}^{2} ]=0$. The bath spin $\hat{J}^{2}$ is conserved quantity only  for homogeneous case, but  not for inhomogeneous case.

When the coupling is homogeneous, the expectation value  of the operator $\langle \hat{J}^{2} \rangle=\frac{N}{2}(\frac{N}{2}+1)$, namely the bath can be regarded as a collective large spin, so that $\tilde{R}=1$. While coupling is inhomogeneous, the angular momentum of the bath  will change with the time. The value of $\tilde{R}$ is determined  by the interaction distribution. The evolution of  bath inhomogeneity $\tilde{R}$ evolve with  time for different inhomogeneity factor $\alpha$, shown in Fig.~\ref{inJ2}.

{\bf The collective bath projector operator $\hat{P}_{bath}$.}\\
The projector operator $\hat{P}_{bath}$ of collective bath spin,  $\hat{P}_{bath}=\sum_{n} \vert n \rangle \langle n \vert$.  The expectation is given
\begin{equation}
\langle \psi(t) \vert \hat{P}_{bath}  \vert \psi(t) \rangle=\sum_{n} \vert \langle n \vert \psi(t) \rangle \vert^{2}.
\end{equation}
For homogeneouse case, we expand the wave function $\vert \psi(t) \rangle$ in the state basises $ \vert \uparrow \rangle_{0}  \vert n \rangle $ and $\vert \downarrow \rangle_{0}  \vert n \rangle$
\begin{equation}
\vert \psi(t) \rangle= \sum_{n}a_{n}(t) \vert \uparrow \rangle_{0}  \vert n \rangle+ \sum_{n}b_{n}(t) \vert \downarrow \rangle_{0}  \vert n \rangle.
\end{equation}
Then we can obtain 
\begin{equation}
\vert \langle n \vert \psi(t) \rangle \vert^2=\vert a_{n}\vert ^{2}+\vert b_{n}\vert^{2}.
\end{equation}
The expectation value  of the projector
\begin{equation}
\langle \psi(t) \vert \hat{P}_{bath}  \vert \psi(t) \rangle=\sum_{n}( \vert a_{n}\vert ^2 +\vert  b_{n} \vert ^2 ).
\end{equation}
Meanwhile,  the normalization
\begin{equation}
\langle \psi(t) \vert \psi(t) \rangle=\sum_{n}( \vert a_{n}\vert ^2 +\vert  b_{n} \vert ^2 )=1
\end{equation}
The expectation of the projector  $\hat{P}_{bath}=\sum_{n} \vert n \rangle \langle n \vert$ equal to one, since the bath projector operator is equivalent to identity operator of homogeneous central spin.
\begin{equation}
I=( \vert \uparrow \rangle   \langle  \uparrow \vert_{0}+\vert \downarrow \rangle   \langle  \downarrow \vert_{0}) \otimes \hat{P}_{bath}
\end{equation}
The expectation value  $\langle \psi(t) \vert I  \vert \psi(t) \rangle=\langle \psi(t) \vert \hat{P}_{bath}   \vert \psi(t) \rangle =1$ always holds, also see  the Fig.4 in the paper,
where we discussed  the expectation value of the bath projector  $\hat{P}_{bath}$  for different inhomogeneous couplings of the  $\alpha$. A large  inhomogeneity factor $\alpha$ makes the wave function have smaller overlap with the collective  state basis  $ \vert \uparrow \rangle_{0}  \vert n \rangle $ and $\vert \downarrow \rangle_{0}  \vert n \rangle$.

\begin{figure}
  \begin{center}
\includegraphics[scale=0.35]{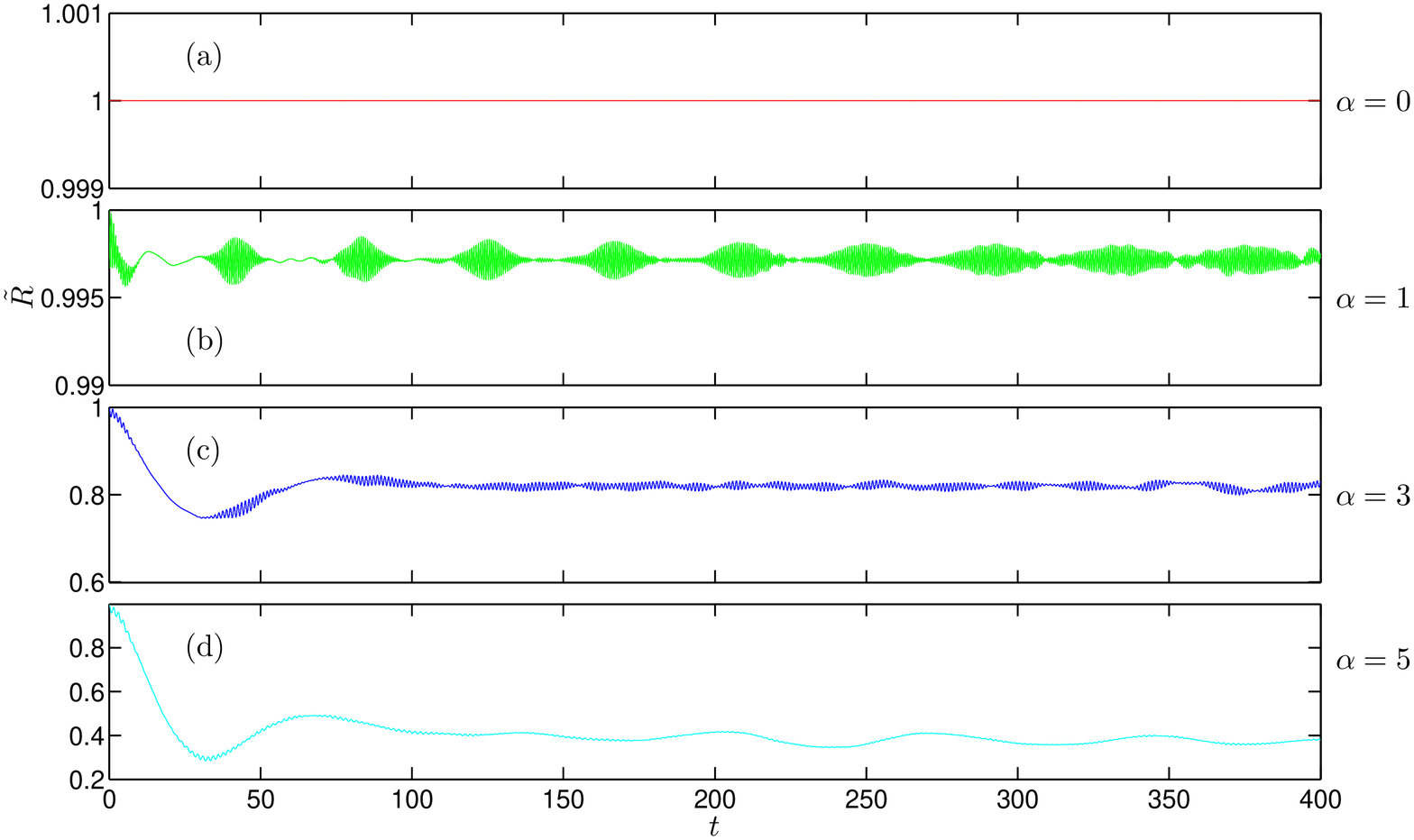}
\end{center} 
\caption{The reduced bath angular momentum evolve with time at different inhomogeneous factor $\alpha
$.  Here we take a setting $N=10, B=A=1$.}
\label{inJ2}
\end{figure}

\begin{figure}
  \begin{center}
\includegraphics[scale=0.4]{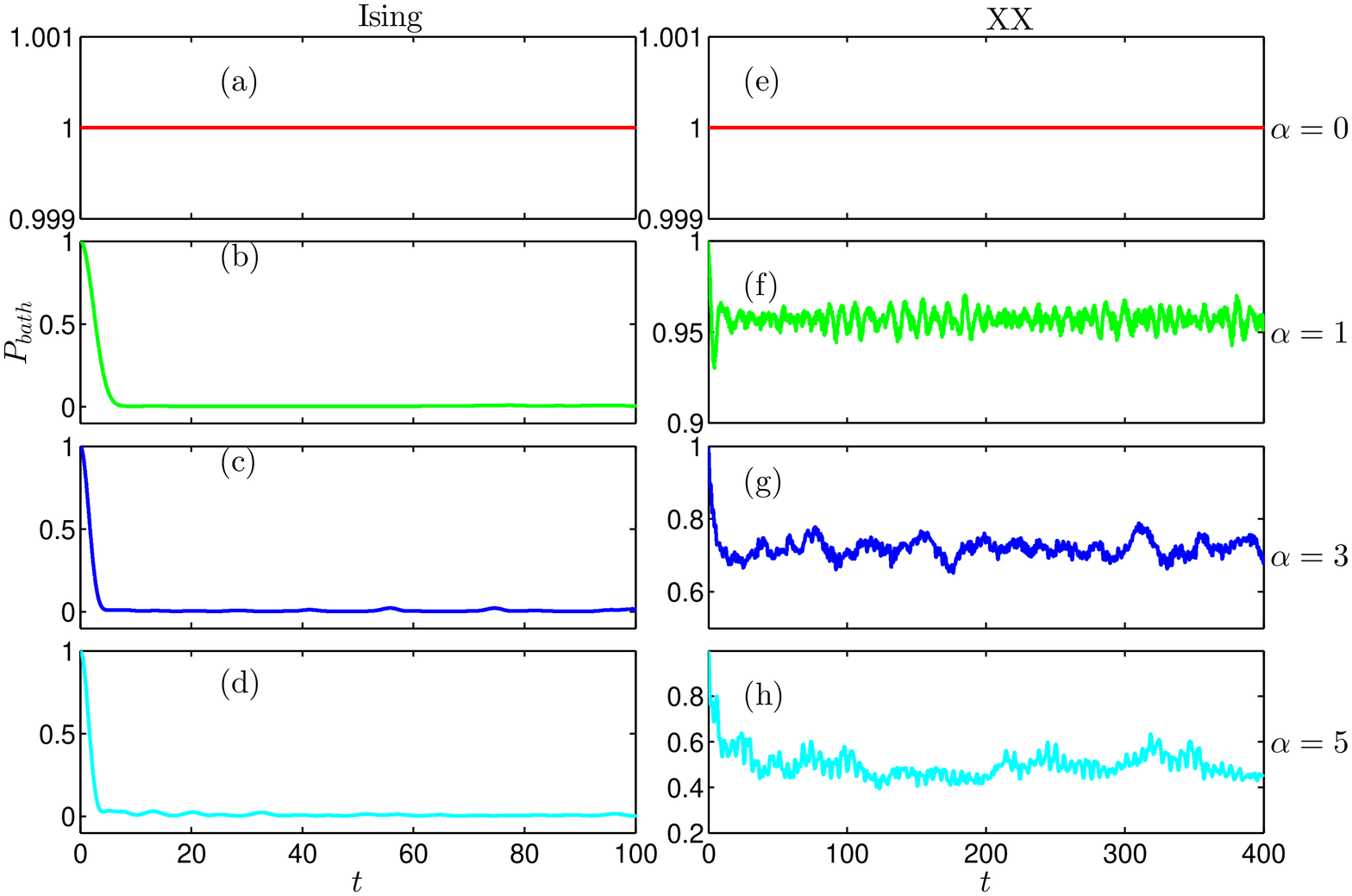}
\end{center} 
\caption{The  the expectation values  of the bath projector for two types of coupling constants:  Ising-type central spin model $H=B\mathbf{\bf s} ^{z}_{0}+2\sum_{j=1}^{N}\Delta_j \,\mathbf{\bf s}_{0}^{z} \mathbf{\bf s}_{j}^{z}$ (left column) with  $\Delta_{j}=A \exp(-\alpha(j-1)/N)$ and XX-type central spin model  $H=B\mathbf{\bf s} ^{z}_{0}+2\sum_{j=1}^{N}A_j( \mathbf{\bf s}_{0}^{x} \mathbf{\bf s}_{j}^{x}+\mathbf{\bf s}_{0}^{y} \mathbf{\bf s}_{j}^{y})$ (right column)  with $A_j=A \exp(-\alpha(j-1)/N)$.  Here we take a setting $N=10, B=A=1$.}
\label{Proj_xxdel}
\end{figure}
}
%\begin{figure}
%  \begin{center}
%\includegraphics[scale=0.4]{Qcrsm_XX&del.eps}
%\end{center} 
%\caption{The  the expectation of the bath projector for two type of coupling,  Ising-type central spin model $H=B\mathbf{\bf s} ^{z}_{0}+2\sum_{j=1}^{N}\Delta_j \,\mathbf{\bf s}_{0}^{z} \mathbf{\bf s}_{j}^{z}$ (left column) and XX-type central spin model  $H=B\mathbf{\bf s} ^{z}_{0}+2\sum_{j=1}^{N}A_j( \mathbf{\bf s}_{0}^{x} \mathbf{\bf s}_{j}^{x}+\mathbf{\bf s}_{0}^{y} \mathbf{\bf s}_{j}^{y})$ (right column) .  here $N=10, B=A=1$.}
%\label{Proj_xxdel}
%\end{figure} 

\end{widetext}


\begin{thebibliography}{99}
\bibitem{note-1} { The Zeeman terms of the central spin and the bath spins  are in general given by $Bs_{0}^{z}+B^{\prime}\sum_{i=1}^NS_i^z=(B-B^{\prime})s_{0}^{z}+B^{\prime}(s_{0}^{z}+\sum_{i=1}^NS_i^z)$. 
In the Hamiltonian (\ref{Ham-1}) and (\ref{Ham-2}), we dropped the the total Zeeman energy term and thus the effective magnetic field of the central spin   was shifted by $(B-B^{\prime})$. }
\end{thebibliography}
\end{document}